\documentclass[9pt,twocolumn,twoside,]{pnas-new}

\templatetype{pnasresearcharticle} 

\title{A backward-spinning star with two coplanar planets}

\author[a,1]{Maria Hjorth}
\author[a,1]{Simon Albrecht} 
\author[b]{Teruyuki Hirano}
\author[c]{Joshua N.\ Winn}
\author[d]{Rebekah I.\ Dawson}
\author[e]{J.\ J.\ Zanazzi}
\author[a]{Emil Knudstrup}
\author[b]{Bun'ei Sato}

\affil[a]{Stellar Astrophysics Centre, Department of Physics and Astronomy, Aarhus University, Ny Munkegade 120, DK-8000 Aarhus C, Denmark}
\affil[b]{Department of Earth and Planetary Sciences, Tokyo Institute of Technology, 2-12-1 Ookayama, Meguro-ku, Tokyo 152-8551, Japan}
\affil[c]{Department of Astrophysical Sciences, Princeton University, 4 Ivy Lane, Princeton, NJ 08544, USA}
\affil[d]{Department of Astronomy \& Astrophysics, Center for Exoplanets and Habitable Worlds, The Pennsylvania State University, University Park, PA 16802, USA}
\affil[e]{Canadian Institute for Theoretical Astrophysics, University of Toronto, 60 St.\ George Street, Toronto, Ontario, M5S 3H8, Canada}

\leadauthor{Hjorth} 

\significancestatement{
The Sun's equator lines up with the orbits of the planets. This fact supports
the theory that stars and their planets inherit their angular momentum from the
same source: the gravitational collapse of a molecular cloud. Most astronomers
expected spin-orbit alignment to be a universal feature of planetary systems.
This proved false: many drastic misalignments are known, and many possible
reasons have been offered. In one theory, a distant companion star upsets the
alignment at an early stage, while the star is still surrounded by a
protoplanetary disk. Here, the K2-290 system is shown to be the best known
candidate for such a primordial misalignment. The star rotates backwards, and a
companion star with suitable properties has been identified.

}

\authorcontributions{MH and SA devised and coordinated the study of K2-290. MH wrote the initial draft of the manuscript with the support of SA. MH, EK, and SA performed the analysis of the Rossiter-McLaughlin data. The observations were led by MH, SA, and TH, with EK performing the HARPS-N observations, MH and TH performing the HDS observations and the ESPRESSO observations being carried out by ESO employees. MH reduced the ESPRESSO data and TH the HDS data, with BS providing the pipeline for the HDS measurements. JNW analyzed the photometry to obtain the stellar rotation period. RID performed the N-body and secular resonance calculations and simulations related to the post formation processes. JJZ did the calculations and integrations of the primordial misalignment process. MH, SA, TH, JNW, RID, and JJZ discussed and commented on the origin of the retrograde planetary orbits, and contributed to the manuscript. All authors reviewed and commented on the manuscript.}

\authordeclaration{There are no conflicts of interest.}
\equalauthors{\textsuperscript{1} MH (Author One) contributed equally to this work with SA (Author Two).}
\correspondingauthor{\textsuperscript{2}To whom correspondence should be addressed. E-mail: albrecht@phys.au.dk}


\begin{abstract} 

It is widely assumed that a star and its protoplanetary disk are initially aligned, with the stellar equator parallel to the disk plane.  When observations reveal a misalignment between stellar rotation and the orbital motion of a planet, the usual interpretation is that the initial alignment was upset by gravitational perturbations that took place after planet formation.  Most of the previously known misalignments involve isolated hot Jupiters, for which planet-planet scattering or secular effects from a wider-orbiting planet are the leading explanations. In theory, star/disk misalignments can result from turbulence during star formation or the gravitational torque of a wide-orbiting companion star, but no definite examples of this scenario are known.  An ideal example would combine a coplanar system of multiple planets --- ruling out planet-planet scattering or other disruptive post-formation events --- with a backward-rotating star, a condition that is easier to obtain from a primordial misalignment than from post-formation perturbations. There are two previously known examples of a misaligned star in a coplanar multi-planet system, but in neither case has a suitable companion star been identified, nor is the stellar rotation known to be retrograde.  Here, we show that the star K2-290\,A is tilted by $124\pm 6$ degrees compared to the orbits of both of its known planets, and has a wide-orbiting stellar companion that is capable of having tilted the protoplanetary disk.  The system provides the clearest demonstration that stars and protoplanetary disks can become grossly misaligned due to the gravitational torque from a neighbouring star. 

\end{abstract}

\dates{This manuscript was compiled on \today}
\doi{\url{www.pnas.org/cgi/doi/10.1073/pnas.2017418118}}

\begin{document}

\maketitle
\thispagestyle{firststyle}
\ifthenelse{\boolean{shortarticle}}{\ifthenelse{\boolean{singlecolumn}}{\abscontentformatted}{\abscontent}}{}

\dropcap{T}he K2-290 system \cite{Hjorth+2019} consists of three stars. The primary star, K2-290\,A, is a late-type F-star with a mass of $1.19\pm0.07$ solar masses ($M_\odot$) and a radius of $1.51\pm 0.07$~solar radii ($R_\odot$).  The secondary star, K2-290\,B, is an M-dwarf with a projected orbital separation of $113\pm2$~astronomical units (au).  The tertiary star, K2-290\,C, is another M-dwarf located further away, with a projected separation of $2467^{+177}_{-155}$~au.  The primary star harbors two transiting planets.  The inner planet ``b'' has an orbital period of 9.2~days and a radius of $3.06\pm0.16$~Earth radii ($R_{\oplus}$), making it a ``hot sub-Neptune.''  The outer planet ``c'' is a ``warm Jupiter'' with orbital period 48.4~days, radius $11.3\pm 0.6~R_\oplus$, and mass $246\pm 15$ Earth masses ($M_\oplus$).

\section*{Stellar obliquity in the K2-290\,A system}

We performed high-resolution optical spectroscopy of K2-290\,A over time intervals spanning the transits of both planets, as a way of measuring the sky-projected obliquity of the star via the Rossiter-McLaughlin (RM) effect.  The physical basis of this method is that the transiting planet blocks a portion of the stellar photosphere, leading to a distortion in the star's absorption lines. When the planet is in front of the approaching half of the rotating star, it blocks some of the light that would ordinarily be blueshifted due to the Doppler effect associated with stellar rotation.  The absorption lines show a small deficit on the blue side, which can be detected directly or as an apparent redshift of the entire line.  The time evolution of the spectral distortions throughout a transit depends on the sky-projected stellar rotational velocity ($v\sin i_\star$) and the sky projection of the stellar obliquity ($\lambda$).  For example, if the stellar rotation and the planet's orbital motion are aligned, then the planet blocks the blueshifted (approaching) side of the star during the first half of the transit, and the redshifted (receding) side of the star during the second half of the transit. This would be observed as an anomalous redshift followed by a blueshift. 

\subsection*{Planet c} 
\begin{figure*}
\centering
\includegraphics[width=0.49\textwidth]{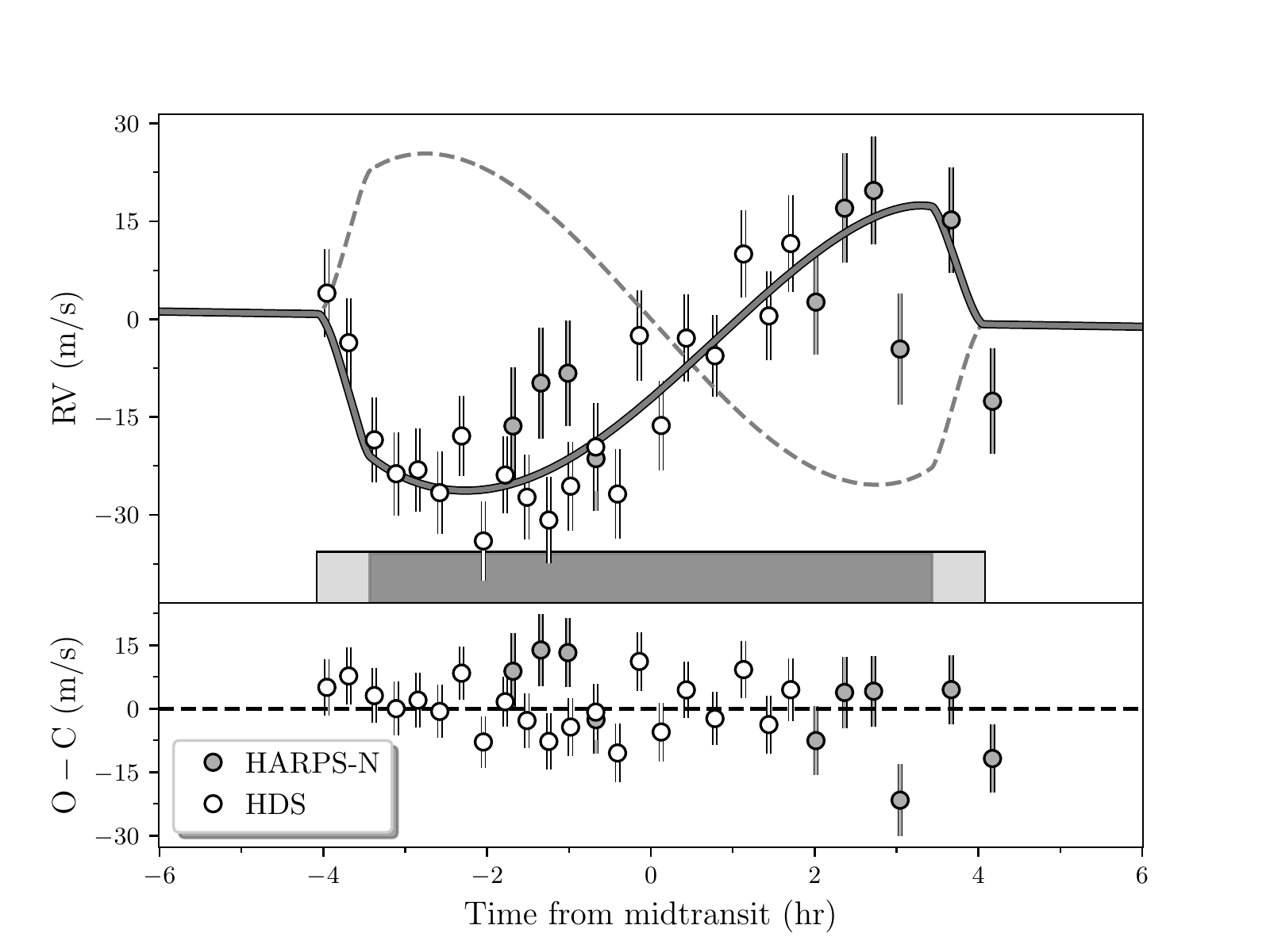}\includegraphics[width=0.49\textwidth]{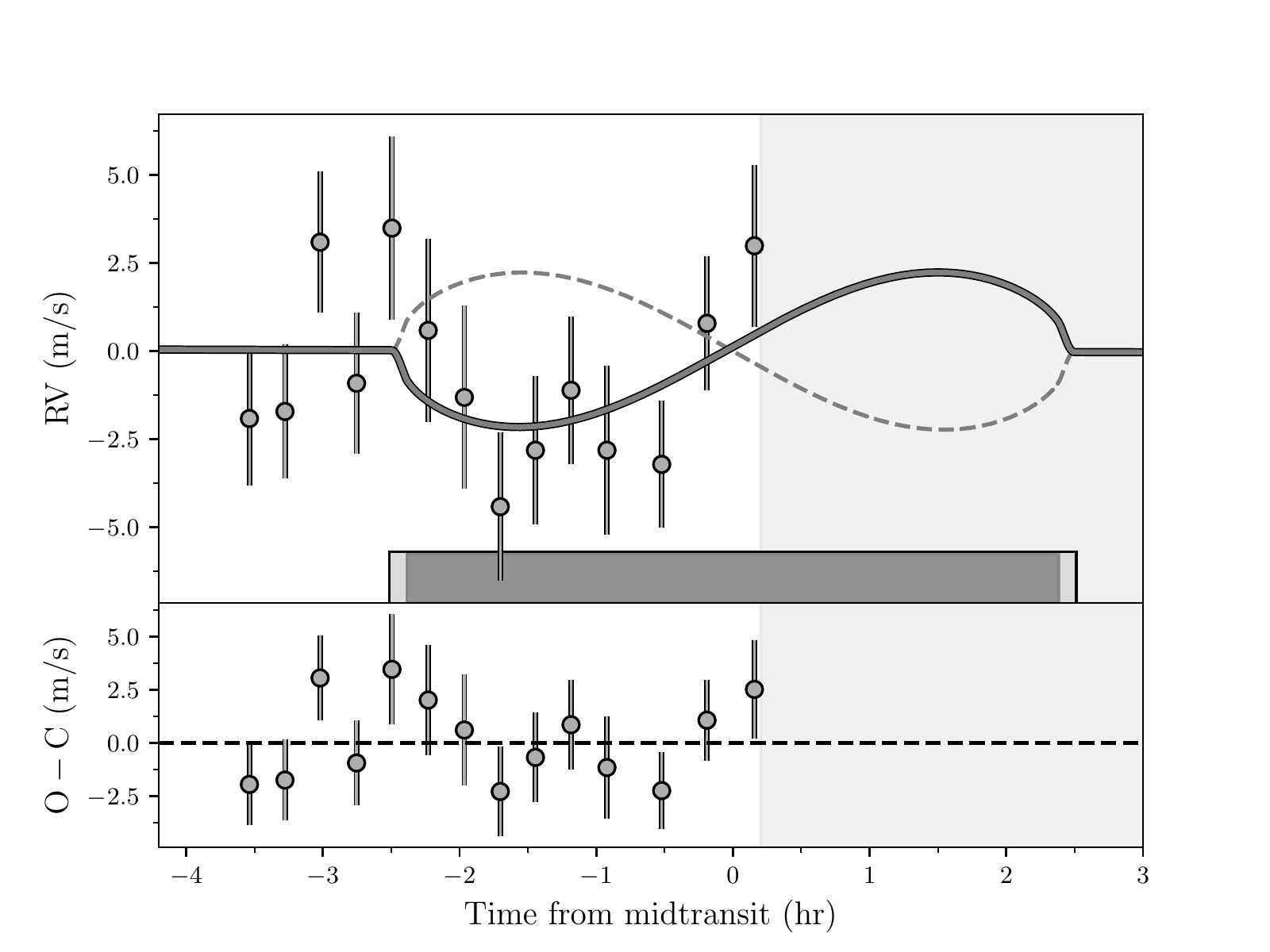} 
\caption{{\bf The Rossiter-McLaughlin effect for both of the planets in the K2-290 system.} Left: Data for the outer, larger planet c. Gray circles are radial-velocity (RV) data obtained with HARPS-N on 25 April 2019. White circles show HDS RVs obtained on 12 June 2019. The error bars indicate the internal uncertainties as derived for the RV data points by the Data Reduction Software of the spectrographs. The solid gray line is the best-fitting model, which has a projected obliquity of $\lambda_{\rm c} = 153 \pm 8$~deg. The lower panel shows Observed minus Calculated (O-C) between the data and the best-fitting model. The dashed gray line is a model in which $\lambda_{\rm c}$ is zero and all the other parameters are the same as in the best-fitting model. The horizontal bars denote the time intervals from first to fourth contact (light gray) and second to third contact (dark gray). Right: Same as left, but for the smaller, inner planet b. These RVs were obtained on 20 July 2019 with ESPRESSO. We find $\lambda_{\rm b} = 173^{+45}_{-53}$~deg. The shaded gray area shows the time interval when our view of K2-290 was blocked by clouds.\label{fig1}}
\end{figure*}

This is the opposite of the pattern that was observed for the outer planet of K2-290 (Fig.~\ref{fig1} left panel). The observed pattern of radial velocities implies that the planet's orbit is retrograde with respect to the star's rotation. The data are based on observations of two different transits.  We observed the first transit on 25 April 2019 with the High Accuracy Radial-velocity Planet Searcher North, [HARPS-N, \cite{Cosentino+2012}] on the 3.6m Telescopio Nazionale Galileo on La Palma, in the Canary Islands.  For the second transit, on 12 June 2019, we used the High Dispersion Spectrograph, [HDS, \cite{Noguchi+2002}] on the 8.2m Subaru Telescope on Mauna Kea, in Hawaii.  Together, these two data sets provide complete phase coverage of the 8.1-hour transit (Supplementary Information). 

By fitting a parameterized model to the RV time series (see Supplementary Information), we find the sky-projected obliquity to be $\lambda_{\rm c} = 153 \pm 8$~deg.  We also find $v\sin i_{\star} = 6.9^{+0.5}_{-0.6}$~km~s$^{-1}$ (Fig.~\ref{fig1}), which is consistent with the value that was determined independently from the observed line-broadening of the stellar spectrum \cite{Hjorth+2019}.

\subsection*{Planet b}
What about the orbital orientation of the inner planet? The observation of the RM effect for the inner planet is more challenging because of the planet's smaller size. One might expect the orbits of the two planets to be closely aligned, based on prior observations and statistical analyses of the {\it Kepler} multi-planet systems   \cite{Fabrycky+2014,FangMargot2012,Zhu+2018}, but K2-290 is unusual in having a giant planet. Furthermore, the statistical studies could not tell whether the planets always orbit in the same direction, or if they could sometimes orbit in opposite directions (an admittedly speculative possibility).

\paragraph*{Limits on mutual orbital inclinations from long-term dynamical simulations}
We performed numerical integrations of the gravitational dynamics of the two-planet system in order to check on the long-term stability of the system for different choices of the mutual orbital inclination. We used the {\tt mercury6} code \cite{Chambers1999}, including the effects of general relativity. We assumed both orbits to be initially circular. This is a conservative assumption in that sense that any initial eccentricity would extend the range of unstable mutual inclinations. The coplanar configuration is stable for at least 2.8~Gyr, the maximum timespan that was simulated. An anti-aligned configuration, with a mutual inclination of $180$~deg, is also stable over the same timespan. 

We tried a configuration in which planet c's orbit is misaligned relative to the star by the observed amount, and planet b is aligned with the stellar equator. This configuration is also stable for at least 2.8~Gyr, although both planets do not often transit at the same time. When one planet transits, the other planet also transits about 15\% of the time.

\begin{figure}
\centering
\includegraphics[width=0.45\textwidth]{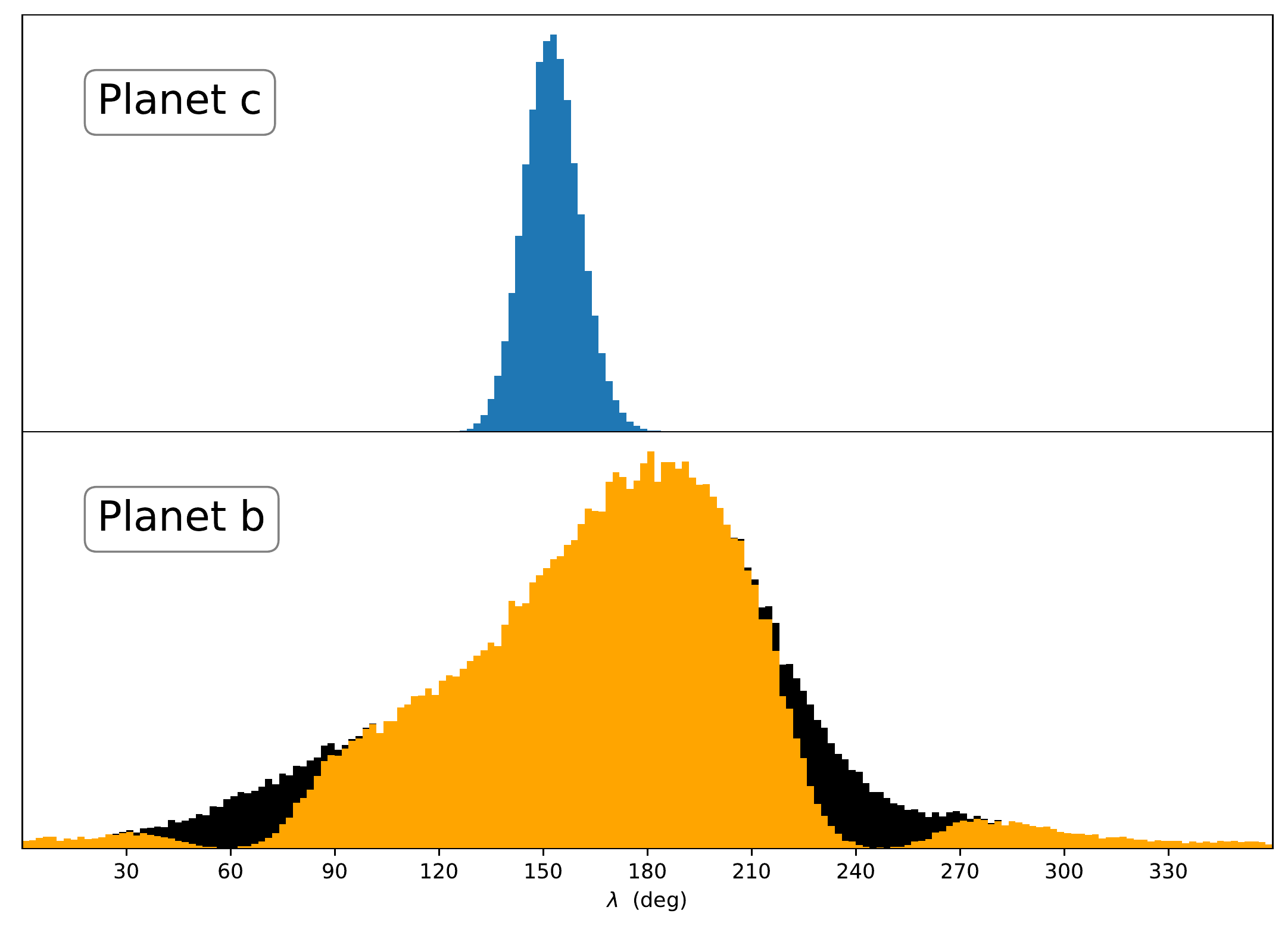} 
\caption{ {\bf Results for the projected obliquities of K2-290 with respect to both of the known planets.} The upper panel shows the posterior for the projected obliquity of planet c, based on the RM measurements shown in Fig.~\ref{fig1}. The lower panel shows the same for the smaller planet b (black). The orange posterior also takes into account the constraints form our  orbital stability calculations.}\label{fig2}
\end{figure}

Based on these integrations, we calculated the expected level of transit-timing variations (TTV) and transit duration variations (TDVs) using the code described in Ref.~\cite{Dawson+2014}. In both scenarios, the expected level of TTVs is too small to be detectable using the currently available data set. The TTVs are on the order of 0.1 minutes for the inner planet, and $0.01$ to $0.1$ minutes for the outer planet (depending on the mass of the inner planet). For the case in which the inner planet is aligned with the stellar spin, the TDVs of the inner planet are on the order of 10 minutes over three years. Our simulations also showed that values of the mutual orbital inclinations between $74^\circ$ and $112^\circ$ are unstable, as Lidov-Kozai cycles of planet b's orbit drive up its eccentricity. This makes the planet prone to collisions with the star, or tidal disruption. We display these limits in Fig.~\ref{fig2}. 

We also tested how different values of the planetary masses and orbital eccentricities influence these stability zones. The most significant influence is from the orbital eccentricity of planet c. For example, taking the eccentricity to be $0.144$, we found that the unstable range of mutual inclinations is enlarged to $59^\circ$ -- $136^\circ$. The best available constraint on the orbital eccentricity based on RV data is a 3-$\sigma$ upper limit of $0.24$ \cite{Hjorth+2019}.

In summary, we found that nearly perpendicular configurations can be ruled out, but mutual inclinations near $0^\circ$ and $180^\circ$ are both viable.  

\paragraph*{Spectroscopic transit observations -  planet b}
To decide between these possibilities, we observed a transit of the inner planet with the newly commissioned Echelle Spectrograph for Rocky Exoplanets and Stable Spectroscopic Observations [ESPRESSO, \cite{Pepe+2010}] and one of the 8.4m Very Large Telescopes at Paranal Observatory, in Chile (Supplementary Information). The data, shown in the right panel of Fig.~\ref{fig1}, right panel, rule out anti-aligned orbits and are consistent with aligned orbits (Fig.~\ref{fig2}). For the determination of the projected obliquity of planet b, we followed the same approach as for planet c, this time setting priors on $v\sin i_{\star}$ and the Kepler-band limb-darkening parameters based on the previous analysis. We used the same constraints on the limb darkening coefficients for the ESPRESSO data that were used for the HARPS-N data.

The result is $\lambda_{\rm b} = 173^{+45}_{-53}$~deg (Supplementary Materials), which is compatible with the value obtained for the outer planet. The lower signal-to-noise ratio means we cannot measure $\lambda_{\rm b}$ as well as $\lambda_{\rm c}$. Indeed, if we did not already have precise prior knowledge of the transit depth, the transit impact parameter, the mid-transit time, and $v\sin i_\star$, we could not have been confident that the RM effect for planet b was detected. Our prior knowledge of those parameters guarantees that the RV anomaly has an amplitude of a few meters per second \cite{albrecht+2011}. The only effectively free parameters in the model are $\lambda_{\rm b}$ and the velocity zeropoint.

For an even simpler test, we fitted the data with two different models: one in which the two planets are exactly aligned ($\lambda_{\rm b} = \lambda_{\rm c}$), and the other in which they are anti-aligned ($\lambda_{\rm b} = 180$~deg~$ +\, \lambda_{\rm c}$). There are no free parameters in either model apart from the velocity zeropoint.  The aligned model has a $\chi^2$ statistic of $13$, while the anti-aligned model is a poorer fit with $\chi^2 = 26$. In both cases, the number of degrees of freedom is 13.  Based on these tests, we conclude that the data favor the aligned case, although they do allow a small probability (on the order of a few percent) for the even more exotic possibility of anti-aligned orbits. 

As a consistency check, we also investigated if the RM effect can be detected as deformations of the stellar lines during the transit (i.e.\ the  Doppler shadow -- see Supplementary Information). We found: (i) The ``planet shadow'' cannot be detected by eye in the time series of cross-correlation functions. (ii) This visual non-detection is consistent with the expected strength of the deformation and the Signal-to-Noise Ratio (SNR) of the data (Fig.~S4, panels A-C). (iii) By stacking all the ESPRESSO observations we detected a signal with the expected characteristics, from which we measured $\lambda_b = 157\pm34^\circ$, consistent with the RV-based result and the expected strength of the RM/planet shadow signal (Fig.~S4, panels D-F). We have more confidence in the analysis of the anomalous RVs, because the method is simpler than the analysis of the Doppler shadow.
    
\subsection*{Stellar rotation period and stellar inclination} 
\begin{figure}
\includegraphics[width=0.45\textwidth]{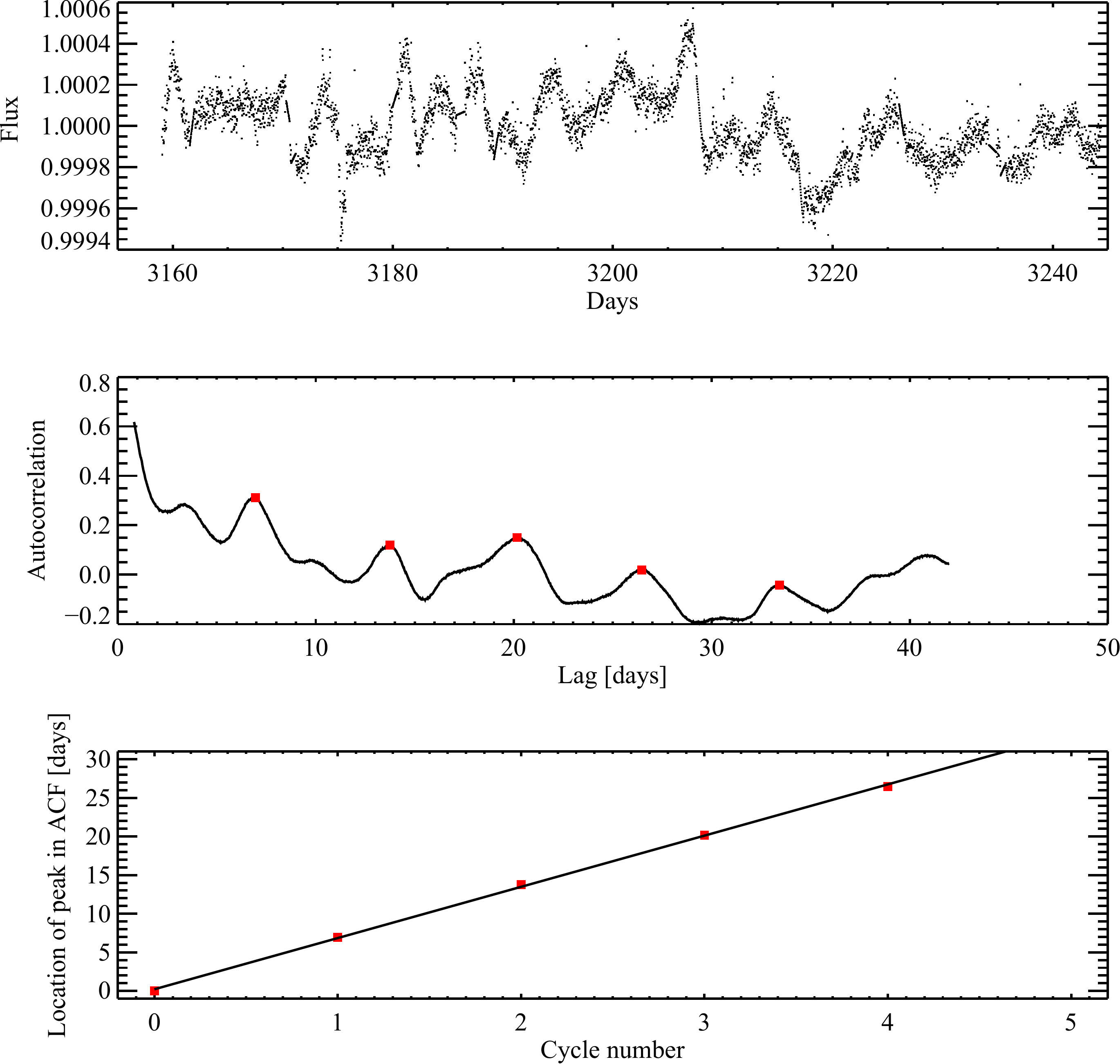}
\caption{\label{fig3} {\bf Determination of the stellar rotation period.} Top: Detrended {\it Kepler} light curve, including linear interpolation across the gaps in the time series. Middle: Autocorrelation as a function of lag. Bottom: Locations of the sequence of autocorrelation peaks in the autocorrelation function (ACF), along with a linear fit. The slope of this line is taken to be the rotation period of K2-290\,A.}
\end{figure}

We determined the stellar rotation period based on the quasi-periodic brightness fluctuations that are seen in the 80-day time series observed with the {\it Kepler} telescope. For this determination, we began with the ``K2SFF'' version of the photometric time series for K2-290\,A \cite{VanderburgJohnson2014}. We removed an overall downward-sloping trend by applying a high-pass median filter with a width of 1{,}000 time samples (20.8 days). To prepare to compute the autocorrelation function, we created a light curve with uniform time sampling by filling in the gaps with linear interpolation.  The resulting light curve is shown in the top panel of Fig.~\ref{fig3}. The middle panel shows the autocorrelation strength as a function of lag, which displays a series of regularly spaced peaks. By fitting a linear function to the measured peak location versus cycle number (bottom panel), we found the slope to be $6.63\pm 0.06$~days.  We adopted this value for the rotation period of K2-290\,A, although we enlarged the fractional uncertainty from 1\% to 10\% to account for systematic errors due to differential rotation \cite{Aigrain+2015,EpsteinPinsonneault2014}.

Based upon this value, we calculated the equatorial rotation velocity $v$ under the assumption of uniform rotation:
\begin{equation}
    v = \frac{2\pi R_\star}{P_{\rm rot}} = 11.2\pm 1.1~{\rm km~s}^{-1},    
\end{equation}
where we have used $R_\star = 1.511\pm 0.075$~$R_\odot$ \cite{Hjorth+2019}. This value for the rotation velocity is larger than the measured projected rotation velocity, $v\sin i_\star = 6.9\pm 0.5$~km~s$^{-1}$, implying that $\sin i_\star$ is less than unity.  Since $\sin i_{\rm orb}$ is known to be close to unity because of the detection of transits, we have here an independent line of evidence for a large stellar obliquity.

We determined the inclination angle of the stellar rotation axis, and the obliquity of the star, $\psi$, using a variation of the Bayesian inference technique advocated by \cite{MasudaWinn2020}. The model parameters were $R_\star$, $P_{\rm rot}$, $\cos i_\star$, for which uniform priors were adopted. The likelihood function was taken to be
\begin{multline}
    {\mathcal L} = 
    \left(\frac{R_\star/R_\odot - 1.511}{0.075}\right)^2 \\
        + \left( \frac{P_{\rm rot} - 6.63\,{\rm d}}{\rm 1\,d} \right)^2
        + \left( \frac{vu - 6.9\,{\rm km/s}}{0.5\,{\rm km/s}} \right)^2,
\end{multline}
where $v \equiv 2\pi R_\star/P_{\rm rot}$ and $u \equiv \sqrt{1-\cos^2 i_\star}$. The result was $\sin i_\star = 0.63\pm 0.09$. Thus we found independent evidence for a large stellar obliquity based on the combination of the stellar rotation period, radius, and projected rotation velocity. There are two possible solutions for $i_\star$, which are near 39$^\circ$ and $141^\circ$.  Combining the posteriors of these two solutions with equal weight, and including the posterior for $\lambda_b$ based on the RM effect, the result for the stellar obliquity is $124\pm 6$~degrees (Table~\ref{tab1}). 

\subsection*{Radial velocity limits on wide-orbiting bodies} By combining two new radial-velocity observations taken with HARPS-N in summer 2019 (the last two entries of Tab.~S1) with the HARPS-N measurements obtained previously \cite{Hjorth+2019}, we have data extending over 500 days. We modeled the planetary signals along with a possible long-term constant acceleration, finding the acceleration to be $\dot{\gamma} = 9\pm 5$~m~s$^{-1}$~yr$^{-1}$. Based on the estimated mass and orbital distance of K2-290\,B ($0.368\pm0.021$~$M_{\odot}$ and $113\pm2$~au), we would expect to see a radial acceleration on the order of 5~m~s$^{-1}$~yr$^{-1}$. Therefore, the observed long-term acceleration is compatible with zero (within 2-$\sigma$) and is also compatible with the expected contribution from the nearby M-dwarf. There is no evidence for any other wide-orbiting bodies.

\section*{Discussion}
\begin{figure}
\centering
\includegraphics[width=0.4\textwidth]{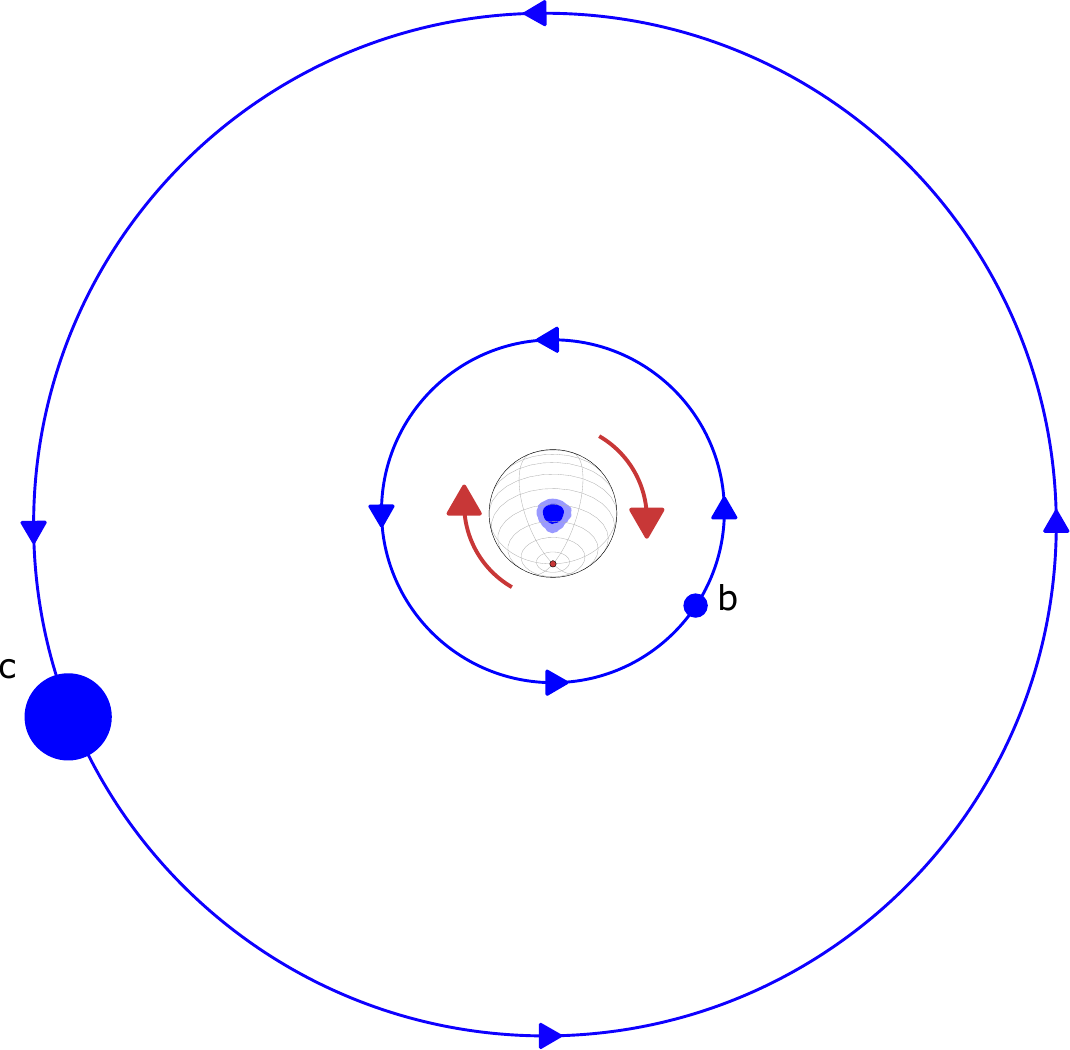} 
\caption{\label{fig4} {\bf Illustration of the architecture of K2-290\,A and its two known planets.}  The view is from above the system's ecliptic plane, along the orbital north pole of the two planets. The ratio of the sizes of the orbits is accurate. Relative to the scale factor of the orbits, the star has been enlarged by a factor of 5, and the planets have been enlarged by a factor of 500. The blue contours indicate the $1$ and $2-\sigma$ confidence intervals for the stellar obliquity,  with respect to planet c. The red point represents the stellar south pole. The red arrows indicate the sense of stellar rotation, and the blue arrows indicate the sense of orbital motion. Our actual view of this system is from the side, allowing for the observation of transits.} 
\end{figure}

Systems similar to K2-290, with coplanar planetary orbits and a grossly misaligned host star (Fig.~\ref{fig4}), had been predicted to exist as a consequence of the tidal torque on a protoplanetary disk from a neighboring star \cite{Batygin2012,BatyginAdams(2013)}. Another possible explanation for such systems is the tilting torque exerted on the inner system of planets by a massive planet on a wide and highly inclined orbit. The Kepler-56 system features two planets on coplanar orbits and a misaligned star \cite{Huber+2013}, and in that case, a wider-orbiting third planet was detected through long-term radial-velocity monitoring \cite{Otor_et_al_2016}.  Based on the mass and orbital distance of the third planet, it is possible or even probable that the planet was responsible for tilting the orbital plane of the inner two planets long after the planets formed \cite{GratiaFabrycky2017}. Likewise, the HD\,3167 multi-planet system was recently found to have a misaligned star \cite{Dalal+2019}, but there is not yet any evidence for either a wider-orbiting planet or a companion star.  Turbulence \cite{Bate+2010} and disk-torquing \cite{Fielding+2015} can lead to misaligned protoplanetary disks. However retrograde orbits, as observed for K2-290A, are difficult to achieve  via turbulence and late infall of material will lead to a further reduction of any misalignment \cite{TakaishiTsukamotoSuto2020}.

The unique aspect of K2-290 is that a companion star has been detected (K2-290\,B) with properties that make it a good candidate for the misalignment of the protoplanetary disk.  There is no evidence for a wider-orbiting massive planet: the upper limit on any long-term radial acceleration is about 10 times smaller than the acceleration that was observed for Kepler-56. In addition, star/disk misalignment is an attractive explanation for K2-290 because it can easily produce retrograde orbits \cite{Lai(2014),ZanazziLai2018}. This is because the orientation of the orbital plane of a wide binary star may have only a weak correlation with the orientation of the protoplanetary disk around either star \cite{JensenAkeson(2014),brinch2016}. In contrast, producing a retrograde system through the action of a wide-orbiting planet requires invoking an unseen third planet with an unusually high orbital inclination which would be difficult to achieve through planet-planet interactions \cite{Chatterjee_et_al_2008}.

\begin{table}
\centering
\caption{\label{tab1} Selected parameters of the K2-290 system. }
\begin{tabular}{lc}
Parameter & Value\\
\midrule
Stellar rotation period, $P_{\rm rot}$ (days) & $6.63\pm0.66$\\
Stellar inclination angle, $i_{\star}$ (deg) & $39 \pm 7$ \\  
Projected stellar rotation velocity, $v\sin i_{\star}$ (km~s$^{-1}$) & $6.9\pm0.5$ \\
Projected obliquity with respect to planet b, $\lambda_{\rm b}$ (deg) & $173^{+45}_{-53}$ \\
Projected obliquity with respect to planet c, $\lambda_{\rm c}$ (deg) & $153 \pm 8$ \\
Obliquity with respect to planet c, $\psi_c$ (deg) & $124\pm 6$ \\
\bottomrule
\end{tabular}
\end{table}

\subsection*{Primordial disk/star misalignment scenario}
The originally proposed mechanism for star/disk misalignment was nodal precession of the disk around the angular momentum vector of the binary orbit \cite{Batygin2012}.  It was later recognized that the gravitational coupling between the star and disk is also important, and that misalignments are more likely to arise from secular resonances between spin and nodal precession \cite{BatyginAdams(2013)}. Furthermore, the magnetic fields of stars as massive as K2-290\,A are probably too weak to have enforced star-disk alignment \citep{SpaldingBatygin2015}.

To demonstrate that this scenario is plausible for the case of K2-290, we calculated the system's secular evolution due to mutual gravitational torques between the rotational bulge of the host star, the protoplanetary disk, and the companion star K2-290\,B. We employed the model described in Ref.\ \cite{ZanazziLai2018}. In this model, there is a star of mass $M_\star$, radius $R_\star$, and rotation frequency $\Omega_\star$. To represent the pre-main sequence phase of stellar evolution, the stellar radius is set equal to $R_\star = 2 \, {\rm R}_\odot$ and the rotation rate is set such that $\bar \Omega_\star = \Omega_\star/\sqrt{G M_\star/R_\star^3} = 0.1$.   Including contraction of the stellar radius, and evolution of the stellar spin during the disk-hosting phase, does not have a significant impact on the ensuing star-disk-binary dynamics \citep{BatyginAdams(2013)}. The star is surrounded by a circular flat disk with an inner radius of $r_{\rm in} = 4\,R_\star$, and an outer radius of $r_{\rm out} = $~50~au. The disk's surface density profile is
\begin{equation}
    \Sigma(r,t) \simeq \frac{M_{\rm d}(t)}{2\pi r_{\rm out} r},
\end{equation}
where
\begin{equation}
    M_{\rm d}(t) = \frac{M_{{\rm d}0}}{1 + t/t_{\rm v}}
\end{equation}
is the disk mass, $t_{\rm v} = 0.5 \, {\rm Myr}$ is the viscous timecale, and $M_{{\rm d}0} = 0.1 \, {\rm M}_\odot$ is the initial disk mass.  Although modifying the disk properties does modify the likelihood of secular resonance crossing, we find a secular resonance occurs over a wide swath of reasonable disk parameters for this system (Supplementary Materials).

The model assumes that the two planets form within the disk at the locations we observe them today. Our model does not take into account the effects of planet migration or photo-ionization of the disk as previous work has shown that these effects tend to lead to even greater excitation of the stellar obliquity during the disk-hosting phase \cite{ZanazziLai2018}. The mass of the outer planet is set equal to its currently observed value, while the mass of planet b is set equal to $21.1 \, {\rm M}_{\oplus}$, the 3-$\sigma$ upper limit that was derived from radial-velocity observations \cite{Hjorth+2019}. In general,  smaller values of $M_{\rm b}$ increase the chance of large star-disk misalignments occurring.  The model also includes a companion star of mass $M_{\rm B}$ in a circular and inclined orbit with radius $a_{\rm B}$. We assumed the binary's semi-major axis is greater than the observed projected separation [$a_{\rm B} > 113 \, {\rm au}$ \cite{Hjorth+2019}]. We used the secular equations from \cite{ZanazziLai2018} for the dynamical evolution of planet-forming star-disk-binary systems. To these, we added the gravitational influence of planets b and c on the star-disk $\tilde \omega_{\rm sd}'$ and disk-star $\tilde \omega_{\rm ds}'$ precession frequencies [see eqs.\ 65-66 of \cite{ZanazziLai2018}]:
 
\begin{equation}
    \tilde{\omega}'_{\rm sd} = \tilde{\omega}_{\rm sd} + \omega_{\rm sb} + \omega_{\rm sc}, 
    \label{eq:tom_sd}
\end{equation}
\begin{equation}
    \tilde{\omega}'_{\rm ds} = \tilde{\omega}_{\rm ds} + (L_{\rm b}/L_{\rm d})\omega_{\rm bs} + (L_{\rm c}/L_{\rm d}) \omega_{\rm cs},
    \label{eq:tom_ds}
\end{equation}
where $L_{\rm d} \simeq (2/3)M_{\rm d} \sqrt{G M_\star r_{\rm out}}$ is the total disk orbital angular momentum, $L_i = M_i \sqrt{G M_\star a_i}$ are the planets' orbital angular momenta (where $i$ is either b or c), and the precession frequencies are
 
\begin{equation}
    \omega_{{\rm s}i} = \frac{3 k_q}{2 k_\star} \bar \Omega_\star \left( \frac{M_i}{M_\star} \right) \frac{\sqrt{G M_\star R_\star^3}}{a_i^3},
\end{equation}
\begin{equation}
    \omega_{i{\rm s}} = \frac{3 k_q}{2} \bar \Omega_\star^2 \left( \frac{R_\star}{a_i} \right)^2 \sqrt{ \frac{G M_\star}{a_i^3} }.
\end{equation}
 In \eqref{eq:tom_sd} and \eqref{eq:tom_ds}, we use the notation of \cite{ZanazziLai2018}, where precession frequencies with tildes are averaged (integrated) over the radial extent of the disk. We assumed the primary star to have moment-of-inertia constants of $k_\star = 0.2$ overall and $k_q = 0.1$ for the rotational bulge, as appropriate for the pre-main sequence phase \cite{Lai(2014)}. We neglected the torque on the planets from star B ($\tilde \omega_{\rm dB}' \simeq \tilde \omega_{\rm dB}$).

\begin{figure}
\includegraphics[width=0.45\textwidth]{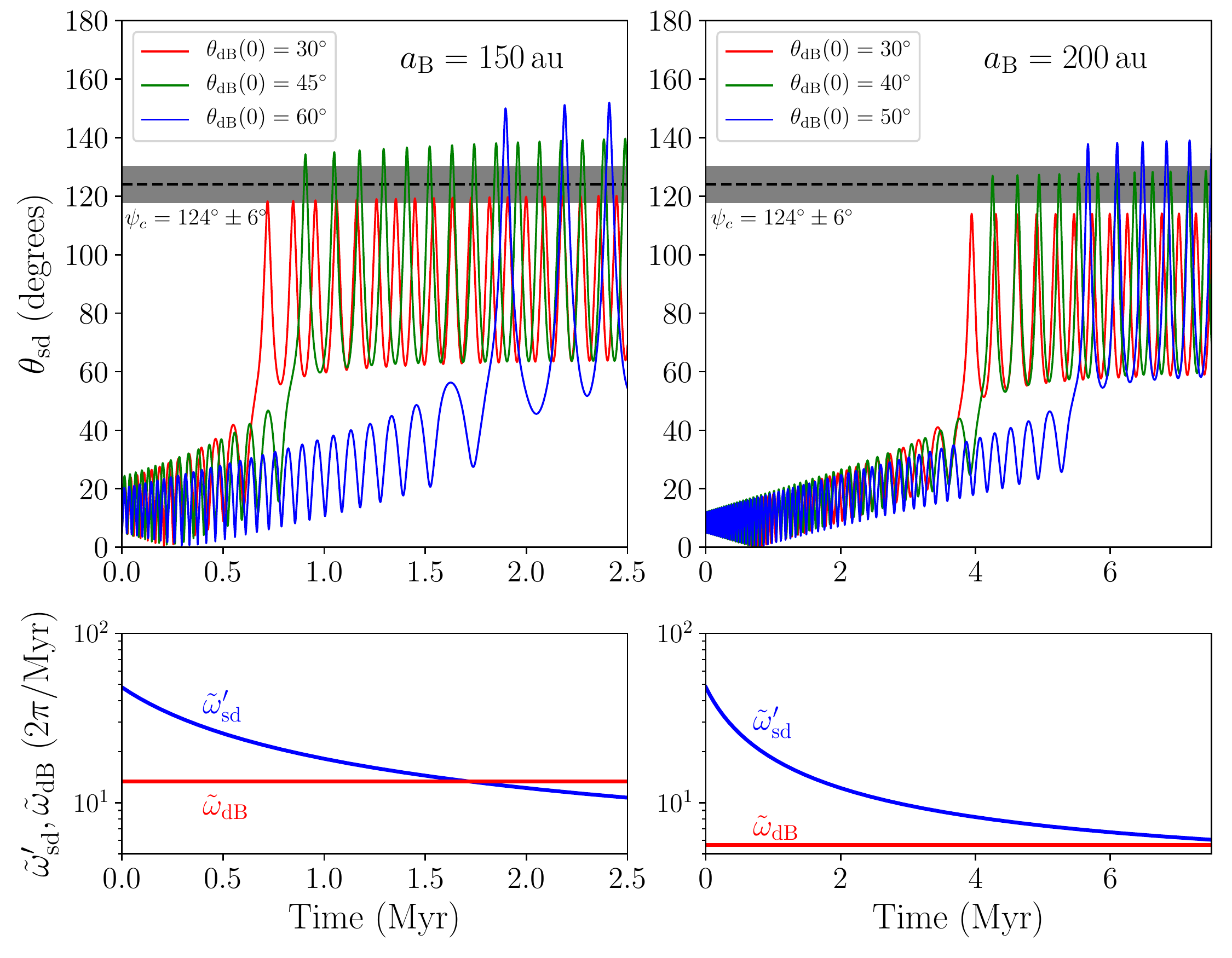}
\caption {\label{fig5}{\bf Simulation of primordial misalignment from star-disk-binary interactions (with embedded planets).} The top panels show the time evolution of the angle between the angular momentum vectors of the host star and the disk ($\theta_{\rm sd}$), for different values of the initial disk-binary inclination $\theta_{\rm dB}(0)$ and binary semi-major axis ($a_{\rm B}$), as indicated in the legend. In all 3 cases shown, $\theta_{\rm sd}$ periodically crosses the interval corresponding to the observed stellar obliquity (dashed black line with gray band representing the uncertainty). The bottom panel shows the evolution of the star-disk $\tilde \omega_{\rm sd}'$ and disk-binary $\tilde \omega_{\rm dB}$ precession frequencies.  Large primordial misalignments $\theta_{\rm sd}$ are excited when secular resonance occurs in the protostellar system ($\tilde \omega_{\rm sd}' \sim \tilde \omega_{\rm dB}$).}
\end{figure}

Fig.~\ref{fig5} displays the main result: the time evolution and excitation of primordial misalignments or mutual star-disk inclinations $\theta_{\rm sd} = \cos^{-1}(\hat s \cdot {\hat l}_{\rm d})$ excited by the binary companion, for different initial disk-binary mutual inclinations $\theta_{\rm db} = \cos^{-1}[{\hat l}_{\rm d}(0) \cdot {\hat l}_{\rm B}]$ (where $\hat s$ is the host star's stellar spin axis, ${\hat l}_{\rm d}$ is the disk's orbital angular momentum unit vector, ${\hat l}_{\rm B}$ is the binary's orbital angular momentum unit vector) and binary semi-major axis values $a_{\rm B}$.  After a secular resonance occurs in the system ($\tilde \omega_{\rm sd}' \sim \tilde \omega_{\rm dB}$), large primordial misalignments are generated, which are consistent with the measured stellar obliquity $\psi_c = 124 \pm 6$~deg over a range of parameter values.  Since we took the inner planet's mass to be the 3-$\sigma$ upper limit obtained from RV measurements \cite{Hjorth+2019}, the suppression of primordial misalignments by short-period, massive planets would not occur for this system \cite{ZanazziLai2018}. A misalignment would only be averted if the disk dissipation timescale were extremely fast ($<0.05$~Myr) or if the disk were extremely compact [$<3$~au, \cite{ZanazziLai2018}], and such disks have to our knowledge not been observed.

For a wide range of possibilities for the angle between the disk and the binary orbit, and for a plausible range of values for the binary orbital distance, the final value of the stellar obliquity can be made consistent with the measurements (Fig.~\ref{fig5}). The main requirement for producing a large star/disk misalignment is the occurrence of a secular resonance. In our model, a one-to-one resonance occurs between the precession frequency of the stellar spin axis around the disk's angular momentum vector, and the precession frequency of the disk's angular momentum vector around the angular momentum of the binary orbit. Previous work found that secular resonance crossing occurs as long as the disk loses its mass over  timescales between 0.05 and 1~Myr \citep{BatyginAdams(2013),ZanazziLai2018}, consistent with measurements of protostellar disk masses and host star accretion rates in the Lupus cluster \cite{Manara+2016}. 

In a subset of the models we considered, the companion star continues to reorient the orbital plane of the two planets after the gaseous disk disappears (Supplementary Materials), through nodal precession or another resonance crossing, but in all such cases the original misalignment occurs while the gaseous disk is still present (Fig.~S7). We also note that while K2-290 has retained a stellar companion, many systems lose their companions after a few Myr \cite{DucheneKraus2013}. Therefore the disk-torquing mechanism described here might have operated even in systems which are now observed as single stars.
   
In short, we find the primordial disk/star misalignment scenario to provide a plausible explanation for the observed properties of K2-290. It will be interesting to perform similar measurements of other systems to see how frequently this scenario occurs,
keeping in mind that there may be more than one way to misalign a disk. Studies of directly imaged disks have not yet turned up any examples of a misaligned star however, there are hints, albeit limited, from \cite{Davies2019} who found evidence that 1/3 of their sample had non-aligned configurations. There are some cases in which the inner and outer disk are misaligned, and obviously the star must be misaligned with one of them \citep{Kraus+2020}. For now, the architecture of the K2-290 system shows that we cannot safely assume that stars and their protoplanetary disks are always well aligned. This may lead to a re-interpretation of other systems with observed misalignments, especially the hot Jupiters \cite{Albrecht+2012}. It also helps to set expectations for future explorations of exoplanetary systems with wider-orbiting planets, more similar to the Solar System.

\matmethods{

 In this article we analyze high resolution spectroscopic data obtained for the K2-290 system during transits of two planets in front of their host star. These observations are detailed in the SI, section 1. We use the apparent RVs, derived from these observations, to determine the projected obliquity of K2-290. We detail our choice of model, parameters and prior information in section 2 of the SI. Here we also describe tests to confirm that our results do not critically depend on the exact RV model setup and prior information supplied. To supplement these results further we have employed alternative analysis techniques for the spectroscopic data (sections 3 \& 4 of the SI). These techniques do not make use of the anomalous RVs observed during transit, rather they rely on information contained in the deformation of the stellar line shapes during planetary transits. Our analysis of these deformations is consistent with our earlier analysis of the RVs. Section 5, the last section of the SI, details the setup of our calculations which explore possible post-formation secular resonance scenarios.

The HARPS-N data reported in this paper are archived in the INAF Science Archive ({\tt https://www.ia2.inaf.it/}) under program ID A39TAC\_2.
The HDS data reported in this paper are archived in the SMOKA Archive ({\tt  https://smoka.nao.ac.jp}) under program ID S19A122.
The ESPRESSO data reported in this paper are archived in the ESO Science Archive ({\tt http://archive.eso.org}) under program ID 2103.C-5041(A).
The Kepler light curve of the K2-290 system from the {\it K2} mission reported in this paper is archived in Mikulski Archive for Space Telescopes, MAST ({\tt https://archive.stsci.edu/}).}

\showmatmethods{} 

\acknow{SA, MH, and EK acknowledge the support from the Danish Council for Independent Research through the DFF Sapere Aude Starting Grant No. 4181-00487B, and the Stellar Astrophysics Centre which funding is provided by The Danish National Research Foundation (Grant agreement no.: DNRF106). This work was supported by JSPS KAKENHI Grant Numbers 16K17660 and 19K14783. Work by JNW was supported by the Heising-Simons Foundation and NASA Award 80NSSC18K1009. RID is supported in part by NASA XRP NNX16AB50G. The authors offer sincere thanks to Akito Tajitsu and Sanghee Lee for assisting with the Subaru observations. The data analyzed in this paper were obtained with the Italian Telescopio Nazionale Galileo (TNG) operated on the island of La Palma by the Fundaci{\'o}n Galileo Galilei of the INAF (Istituto Nazionale di Astrofisica) at the Spanish Observatorio del Roque de los Muchachos of the Instituto de Astrofisica de Canarias as part of the TAC programme A39TAC\_2; the Subaru Telescope, which is operated by the National Astronomical Observatory of Japan as part of the programme S19A122; and the Very Large Telescope (VLT) with data collected at the European Organisation for Astronomical Research in the Southern Hemisphere under ESO DDT programme 2103.C-5041(A). The authors wish to recognize and acknowledge the very significant cultural role and reverence that the summit of Maunakea has always had within the indigenous Hawaiian community.  We are most fortunate to have the opportunity to conduct observations from this mountain.}  This paper also includes data collected by the \textit{K2} mission, which was funded by the NASA Science Mission directorate. This research made use of Lightkurve, a Python package for Kepler and TESS data analysis (Lightkurve Collaboration, 2018).

\showacknow{} 



\begin{thebibliography}{10}

\bibitem{Hjorth+2019}
M {Hjorth}, et~al., {K2-290: a warm Jupiter and a mini-Neptune in a triple-star
  system}.
\newblock {\em\protect\JournalTitle{Monthly Notices of the Royal Astronomical
  Society}} \textbf{484}, 3522--3536 (2019).

\bibitem{Cosentino+2012}
R {Cosentino}, et~al., {Harps-N: the new planet hunter at TNG} in {\em
  Ground-based and Airborne Instrumentation for Astronomy IV}, Society of
  Photo-Optical Instrumentation Engineers (SPIE) Conference Series.
\newblock Vol.{} 8446, p. 84461V (2012).

\bibitem{Noguchi+2002}
K {Noguchi}, et~al., {High Dispersion Spectrograph (HDS) for the Subaru
  Telescope}.
\newblock {\em\protect\JournalTitle{Publications of the ASJ}} \textbf{54},
  855--864 (2002).

\bibitem{Fabrycky+2014}
DC {Fabrycky}, et~al., {Architecture of Kepler's Multi-transiting Systems. II.
  New Investigations with Twice as Many Candidates}.
\newblock {\em\protect\JournalTitle{The Astrophysical Journal}} \textbf{790},
  146 (2014).

\bibitem{FangMargot2012}
J {Fang}, JL {Margot}, {Architecture of Planetary Systems Based on Kepler Data:
  Number of Planets and Coplanarity}.
\newblock {\em\protect\JournalTitle{The Astrophysical Journal}} \textbf{761},
  92 (2012).

\bibitem{Zhu+2018}
W {Zhu}, C {Petrovich}, Y {Wu}, S {Dong}, J {Xie}, {About 30\% of Sun-like
  Stars Have Kepler-like Planetary Systems: A Study of Their Intrinsic
  Architecture}.
\newblock {\em\protect\JournalTitle{The Astrophysical Journal}} \textbf{860},
  101 (2018).

\bibitem{Chambers1999}
JE {Chambers}, {A hybrid symplectic integrator that permits close encounters
  between massive bodies}.
\newblock {\em\protect\JournalTitle{Monthly Notices of the Royal Astronomical
  Society}} \textbf{304}, 793--799 (1999).

\bibitem{Dawson+2014}
RI {Dawson}, et~al., {Large Eccentricity, Low Mutual Inclination: The
  Three-dimensional Architecture of a Hierarchical System of Giant Planets}.
\newblock {\em\protect\JournalTitle{The Astrophysical Journal}} \textbf{791},
  89 (2014).

\bibitem{Pepe+2010}
FA {Pepe}, et~al., {ESPRESSO: the Echelle spectrograph for rocky exoplanets and
  stable spectroscopic observations} in {\em Ground-based and Airborne
  Instrumentation for Astronomy III}, of Proceedings of the SPIE.
\newblock Vol.{} 7735, p. 77350F (2010).

\bibitem{albrecht+2011}
S {Albrecht}, et~al., {Two Upper Limits on the Rossiter-Mclaughlin Effect, with
  Differing Implications: WASP-1 has a High Obliquity and WASP-2 is
  Indeterminate}.
\newblock {\em\protect\JournalTitle{The Astrophysical Journal}} \textbf{738},
  50 (2011).

\bibitem{VanderburgJohnson2014}
A {Vanderburg}, JA {Johnson}, {A Technique for Extracting Highly Precise
  Photometry for the Two-Wheeled Kepler Mission}.
\newblock {\em\protect\JournalTitle{The Publications of the Astronomical
  Society of the Pacific}} \textbf{126}, 948 (2014).

\bibitem{Aigrain+2015}
S {Aigrain}, et~al., {Testing the recovery of stellar rotation signals from
  Kepler light curves using a blind hare-and-hounds exercise}.
\newblock {\em\protect\JournalTitle{Monthly Notices of the Royal Astronomical
  Society}} \textbf{450}, 3211--3226 (2015).

\bibitem{EpsteinPinsonneault2014}
CR {Epstein}, MH {Pinsonneault}, {How Good a Clock is Rotation? The Stellar
  Rotation-Mass-Age Relationship for Old Field Stars}.
\newblock {\em\protect\JournalTitle{The Astrophysical Journal}} \textbf{780},
  159 (2014).

\bibitem{MasudaWinn2020}
K {Masuda}, JN {Winn}, {On the Inference of a Star's Inclination Angle from its
  Rotation Velocity and Projected Rotation Velocity}.
\newblock {\em\protect\JournalTitle{The Astronomical Journal}} \textbf{159}, 81
  (2020).

\bibitem{Batygin2012}
K {Batygin}, {A primordial origin for misalignments between stellar spin axes
  and planetary orbits}.
\newblock {\em\protect\JournalTitle{Nature}} \textbf{491}, 418--420 (2012).

\bibitem{BatyginAdams(2013)}
K {Batygin}, FC {Adams}, {Magnetic and Gravitational Disk-Star Interactions: An
  Interdependence of PMS Stellar Rotation Rates and Spin-Orbit Misalignments}.
\newblock {\em\protect\JournalTitle{The Astrophysical Journal}} \textbf{778},
  169 (2013).

\bibitem{Huber+2013}
D {Huber}, et~al., {Stellar Spin-Orbit Misalignment in a Multiplanet System}.
\newblock {\em\protect\JournalTitle{Science}} \textbf{342}, 331--334 (2013).

\bibitem{Otor_et_al_2016}
OJ {Otor}, et~al., {The Orbit and Mass of the Third Planet in the Kepler-56
  System}.
\newblock {\em\protect\JournalTitle{The Astronomical Journal}} \textbf{152},
  165 (2016).

\bibitem{GratiaFabrycky2017}
P {Gratia}, D {Fabrycky}, {Outer-planet scattering can gently tilt an inner
  planetary system}.
\newblock {\em\protect\JournalTitle{Monthly Notices of the Royal Astronomical
  Society}} \textbf{464}, 1709--1717 (2017).

\bibitem{Dalal+2019}
S {Dalal}, et~al., {Nearly polar orbit of the sub-Neptune HD 3167 c.
  Constraints on the dynamical history of a multi-planet system}.
\newblock {\em\protect\JournalTitle{Astronomy \& Astrophysics}} \textbf{631},
  A28 (2019).

\bibitem{Bate+2010}
MR {Bate}, G {Lodato}, JE {Pringle}, {Chaotic star formation and the alignment
  of stellar rotation with disc and planetary orbital axes}.
\newblock {\em\protect\JournalTitle{Monthly Notices of the Royal Astronomical
  Society}} \textbf{401}, 1505--1513 (2010).

\bibitem{Fielding+2015}
DB {Fielding}, CF {McKee}, A {Socrates}, AJ {Cunningham}, RI {Klein}, {The
  turbulent origin of spin-orbit misalignment in planetary systems}.
\newblock {\em\protect\JournalTitle{Monthly Notices of the Royal Astronomical
  Society}} \textbf{450}, 3306--3318 (2015).

\bibitem{TakaishiTsukamotoSuto2020}
D {Takaishi}, Y {Tsukamoto}, Y {Suto}, {Star-disc alignment in the
  protoplanetary discs: SPH simulation of the collapse of turbulent molecular
  cloud cores}.
\newblock {\em\protect\JournalTitle{Monthly Notices of the Royal Astronomical
  Society}} \textbf{492}, 5641--5654 (2020).

\bibitem{Lai(2014)}
D {Lai}, {Star-disc-binary interactions in protoplanetary disc systems and
  primordial spin-orbit misalignments}.
\newblock {\em\protect\JournalTitle{Monthly Notices of the Royal Astronomical
  Society}} \textbf{440}, 3532--3544 (2014).

\bibitem{ZanazziLai2018}
JJ {Zanazzi}, D {Lai}, {Planet formation in discs with inclined binary
  companions: can primordial spin-orbit misalignment be produced?}
\newblock {\em\protect\JournalTitle{Monthly Notices of the Royal Astronomical
  Society}} \textbf{478}, 835--851 (2018).

\bibitem{JensenAkeson(2014)}
ELN {Jensen}, R {Akeson}, {Misaligned protoplanetary disks in a young binary
  star system}.
\newblock {\em\protect\JournalTitle{Nature}} \textbf{511}, 567--569 (2014).

\bibitem{brinch2016}
C {Brinch}, JK {J{\o}rgensen}, MR {Hogerheijde}, RP {Nelson}, O {Gressel},
  {Misaligned Disks in the Binary Protostar IRS 43}.
\newblock {\em\protect\JournalTitle{The Astrophysical Journal}} \textbf{830},
  L16 (2016).

\bibitem{Chatterjee_et_al_2008}
S {Chatterjee}, EB {Ford}, S {Matsumura}, FA {Rasio}, {Dynamical Outcomes of
  Planet-Planet Scattering}.
\newblock {\em\protect\JournalTitle{The Astrophysical Journal}} \textbf{686},
  580--602 (2008).

\bibitem{SpaldingBatygin2015}
C {Spalding}, K {Batygin}, {Magnetic Origins of the Stellar Mass-Obliquity
  Correlation in Planetary Systems}.
\newblock {\em\protect\JournalTitle{The Astrophysical Journal}} \textbf{811},
  82 (2015).

\bibitem{Manara+2016}
CF {Manara}, et~al., {Evidence for a correlation between mass accretion rates
  onto young stars and the mass of their protoplanetary disks}.
\newblock {\em\protect\JournalTitle{Astronomy \& Astrophysics}} \textbf{591},
  L3 (2016).

\bibitem{DucheneKraus2013}
G {Duch{\^e}ne}, A {Kraus}, {Stellar Multiplicity}.
\newblock {\em\protect\JournalTitle{Annual Review of Astronomy and
  Astrophysics}} \textbf{51}, 269--310 (2013).

\bibitem{Davies2019}
CL {Davies}, {Star-disc (mis-)alignment in Rho Oph and Upper Sco: insights from
  spatially resolved disc systems with K2 rotation periods}.
\newblock {\em\protect\JournalTitle{Monthly Notices of the Royal Astronomical
  Society}} \textbf{484}, 1926--1935 (2019).

\bibitem{Kraus+2020}
S {Kraus}, et~al., {A triple-star system with a misaligned and warped
  circumstellar disk shaped by disk tearing}.
\newblock {\em\protect\JournalTitle{Science}} \textbf{369}, 1233--1238 (2020).

\bibitem{Albrecht+2012}
S {Albrecht}, et~al., {Obliquities of Hot Jupiter Host Stars: Evidence for
  Tidal Interactions and Primordial Misalignments}.
\newblock {\em\protect\JournalTitle{The Astrophysical Journal}} \textbf{757},
  18 (2012).

\end{thebibliography}
\end{document}



\maketitle

\SItext

\section{Spectroscopic transit data}

\subsection{Spectroscopic transit observations of planet c}
We observed the first transit on 25 April 2019 with the High Accuracy Radial-velocity Planet Searcher North, [HARPS-N, \cite{Cosentino+2012}] on the 3.6m Telescopio Nazionale Galileo on La Palma, in the Canary Islands. The observations began just before mid-transit and lasted throughout the second half of the transit. There was a two-hour interruption when the telescope was commandeered by other observers pursuing a Target of Opportunity. The HARPS-N spectra cover the wavelength range from $383$ to $693$~nm, with a spectral resolution of $115{,}000$. The exposure time ranged from 20 to 45~minutes. All HARPS-N observations obtained during the transit night are presented in Tab.~\ref{table:RVs_HARPSN}.

For the second transit, on 12 June 2019, we used the High Dispersion Spectrograph, [HDS, \cite{Noguchi+2002}] on the 8.2m Subaru Telescope on Mauna Kea, in Hawaii. With HDS, we obtained 3 reference spectra of K2-290\,A on 11 June 2019, the night before a transit of the outer planet. These served to establish a radial-velocity baseline.  On the transit night of 12 June 2019, we began observing at ingress and were able to continue for about 3/4 of the transit duration. We also obtained three pre-ingress spectra, but none of them proved useful: the first two spectra were excessively contaminated by twilight, and the third exposure suffered from a problem with the detector readout process. For all of the HDS observations, we used Image Slicer \#2 and the Iodine (I$_2$) absorption cell in the light path. The HDS spectra cover the wavelength range from $494$ to $759$~nm, although the iodine absorption lines are almost all within the range from about 500 to 600~nm.  The spectral resolution was approximately 80{,}000 and the exposure time ranged from 15 to 20~minutes. Rather than investing Subaru observing time to obtain an iodine-free template spectrum, we created a template by co-adding archival HARPS and HARPS-N spectra (which have a higher spectral resolution than our HDS spectra). The HDS RVs are listed in Tab.~\ref{table:RVs_HDS}. Together, these two data sets provide complete phase coverage of the 8.1-hour transit. 

\subsection{Spectroscopic transit observations of planet b}

The observation of a transit of the inner planet was performed on 20 July 2019 with ESPRESSO \cite{Pepe+2010}, which was connected to Unit Telescope 3. The ESPRESSO spectra cover the wavelength range from $380$ to $788$~nm, with a spectral resolution of $140{,}000$.  The exposure time ranged from $15$ to $30$~minutes. We extracted Cross Correlation Functions (CCFs) and RVs using the standard Data Reduction Software.

Observations began prior to the ingress and lasted for most of the transit.  Clouds began covering the sky at around the time of the transit midpoint. The effects were to lower the signal-to-noise ratio of each spectrum, increase the contamination of the spectra by moonlight, and increase the strength of telluric spectral features. We chose not to include the affected spectra in our analysis, although for completeness, Tab.~\ref{table:RVs_ESPRESSO} provides the RVs that were extracted from the affected spectra (marked with a $\star$ symbol).

\section{Details of determining projected obliquities via anomalous transit RVs}

To measure projected obliquities of K2-290\,A with respect to planets b and c, we fitted a parameterized model to time series of apparent radial velocities and the photometric time series obtained with the NASA {\it Kepler} telescope during the {\it K2} mission. The model was based on the one that was described in Ref.~\cite{Hjorth+2019}, extended to include the Rossiter-McLaughlin effect \cite{Hirano+2011}.  The photometric time series was obtained via the Lightkurve tool \cite{Lightkurve2018}, which we used to extract a light curve from the {\it K2} data. The resulting light curve has slightly better noise statistics than the time series used in the discovery paper. The RV time series include observations from the transit nights, as well as the three HDS data points obtained the night before a transit of planet c. 

The model parameters were the projected spin-orbit angles of the two planets ($\lambda_{\rm b}$ and $\lambda_{\rm c}$), the projected stellar rotation velocity ($v\sin i_{\star}$), quadratic limb darkening parameters in the V-band ($u_{1,\rm V}$ and $u_{2,\rm V}$), and terms representing the stellar microturbulent and macroturbulent velocities ($v_{\rm mic}$ and $v_{\rm mac}$). There were also some nuisance parameters relating to the HDS and ESPRESSO spectrographs: the additive radial-velocity offsets (``zeropoints'') to align them with the HARPS-N velocity scale ($\gamma_{\rm HDS}$ and $\gamma_{\rm ESP}$). 

The parameters governed mainly by the photometry were the orbital periods ($P_{\rm b,c}$), the times of midtransit ($T_{\rm b,c}$), the scaled orbital separations ($a_{\rm b,c}/R_\star$), the radius ratios ($r_{\rm b,c}/R_\star$), the orbital inclinations ($i_o{\rm b,c}$) and the quadratic limb darkening parameters ($u_{1,\rm K}$ and $u_{2,\rm K}$) for the {\it Kepler} bandpass. 

We used the best estimate of the stellar density  ($\rho_\star=0.485^{+0.074}_{-0.064}~$g\,cm$^{-3}$) from \cite{Hjorth+2019} as a prior constraint, which sharpened the determination $a_{\rm b,c}/R_\star$ and the orbital inclinations. The results for the projected obliquities do not depend critically on this constraint. Additional prior constraints were placed on the orbital RV semi-amplitude for planet c ($K_c=38.4\pm1.7$~m\,s$^{-1}$) and the $3-\sigma$ upper limit of $K_b<6.6$~m\,s$^{-1}$ on planet b \cite{Hjorth+2019}. 

For determining the projected obliquity of planet c we imposed a Gaussian prior constraint on $v\sin i_{\star}$, with a mean of 6.5~km~s$^{-1}$ and a standard deviation of 1.0~km~s$^{-1}$, based on an analysis of the spectral line widths \cite{Hjorth+2019}. For the limb darkening parameters, we used Gaussian priors based on the tabulated values $u_{1,\rm K}=0.38\pm 0.10$ and $u_{2,\rm K}=0.14\pm 0.10$, $u_{1,\rm HARPS-N}=0.37\pm 0.10$ and $u_{2,\rm HARPS-N}=0.31\pm 0.10$, and $u_{1,\rm HDS}=0.41\pm 0.10$ and $u_{2,\rm HDS}=0.33\pm 0.10$ \cite{EastmanGaudiAgol2013}. For the rest of the parameters, we used uniform priors. 

To calculate the posterior distribution of the parameters, we used the Markov Chain Monte Carlo Method (MCMC) as implemented in the {\tt emcee} code \cite{Foreman-Mackey+2013}. We used 100 walkers in both runs and let the code running until the chains had reached a length of 50 times the autocorrelation time at which point we considered the runs to be converged. For planet c this took 12{,}200 draws and 56{,}300 for planet b. We discarded 2 times the autocorrelation time of the first draws in both runs. 

\begin{figure}
\includegraphics[width=0.5\textwidth]{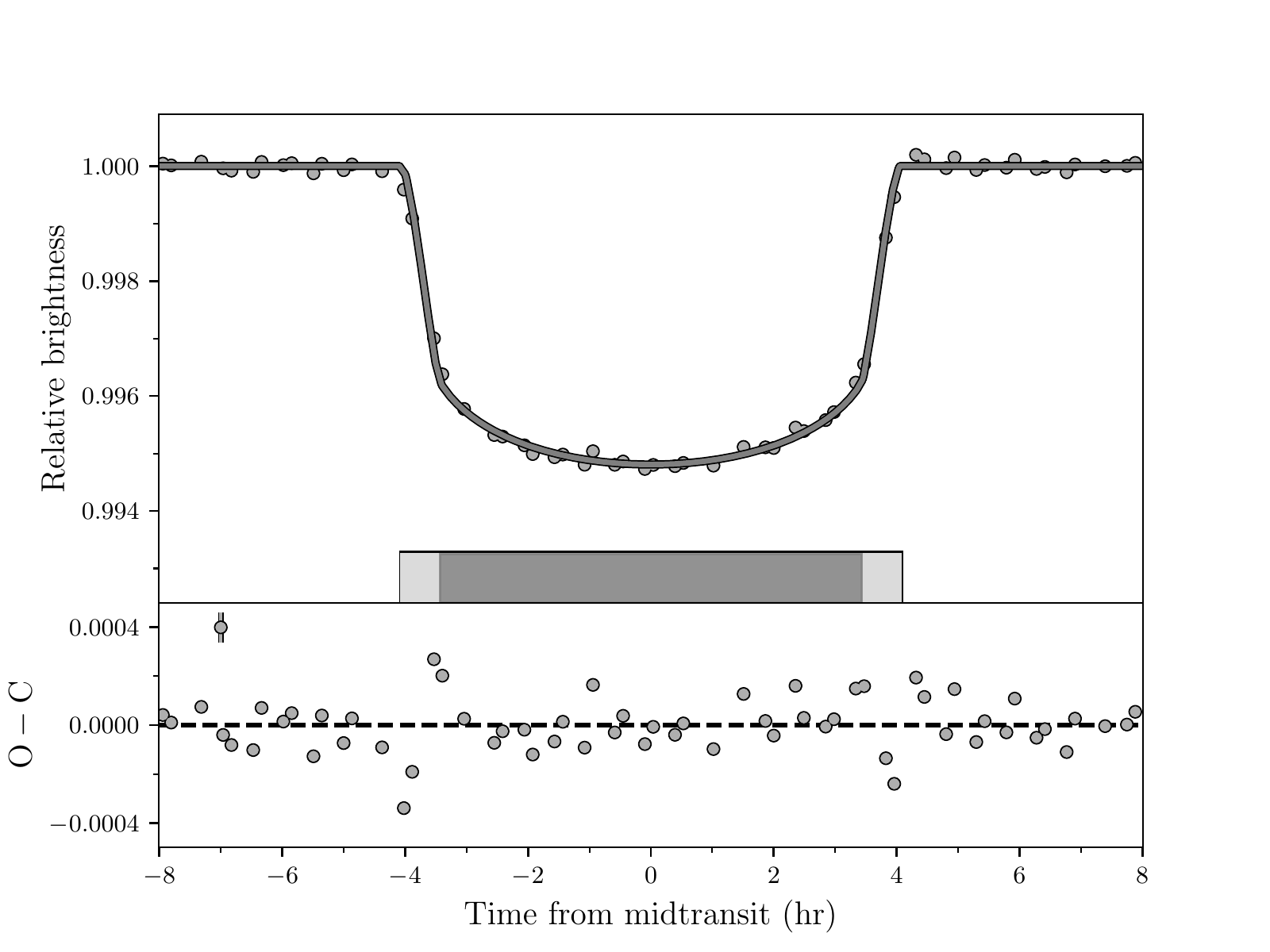}\includegraphics[width=0.5\textwidth]{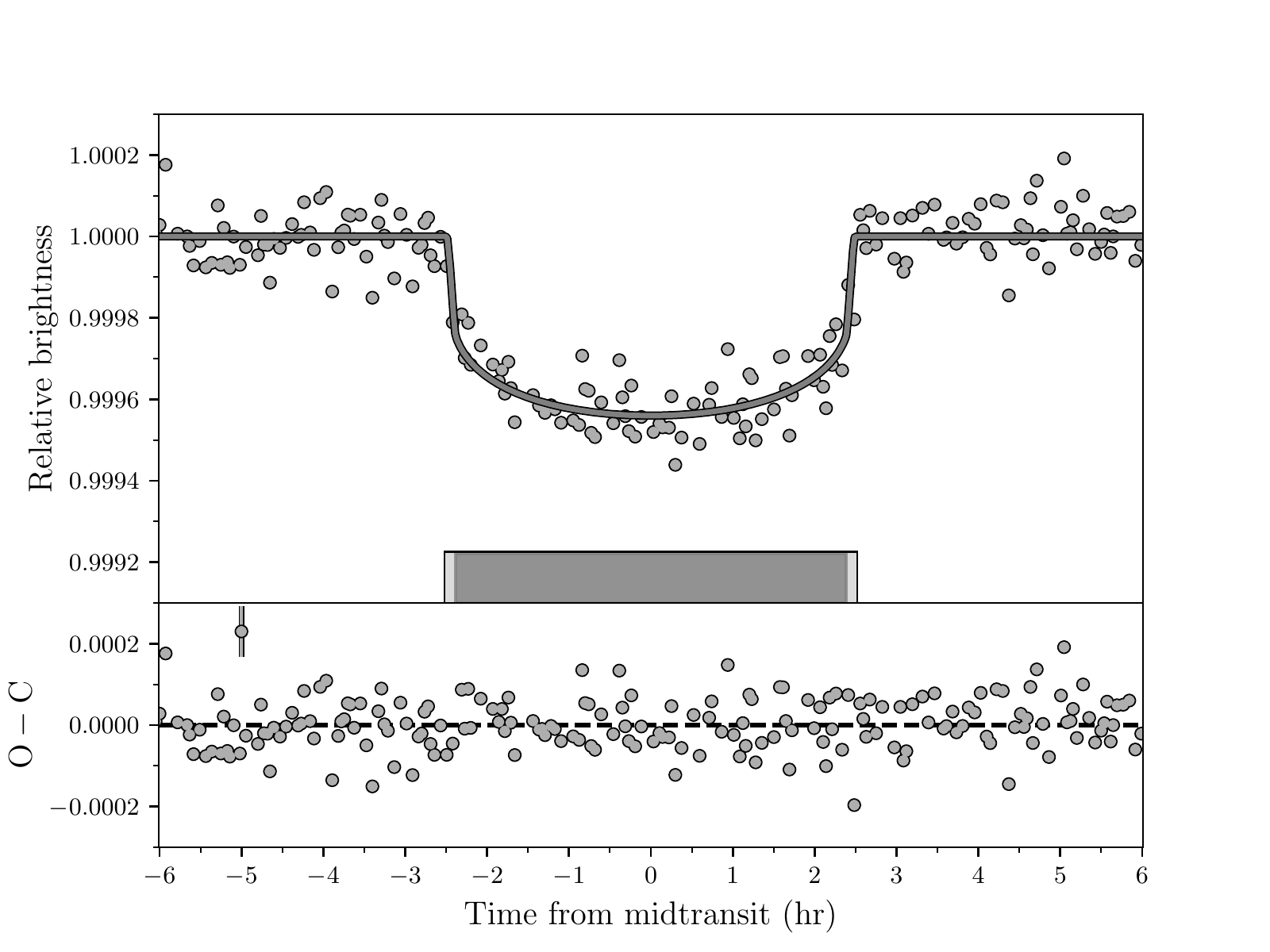} 
\caption{\label{fig:S1}  K2 photometry of K2-290.} Left: Phase-folded light curve of the transits of planet c. The gray line is the best-fitting model. The lower panel displays  Observed minus Calculated (O-C). The single point with an error bar is not a data point; it illustrates the 1-$\sigma$ uncertainty assigned to each data point. Right: Same, but for planet b.
\end{figure}

\begin{figure}
\includegraphics[width=0.5\textwidth]{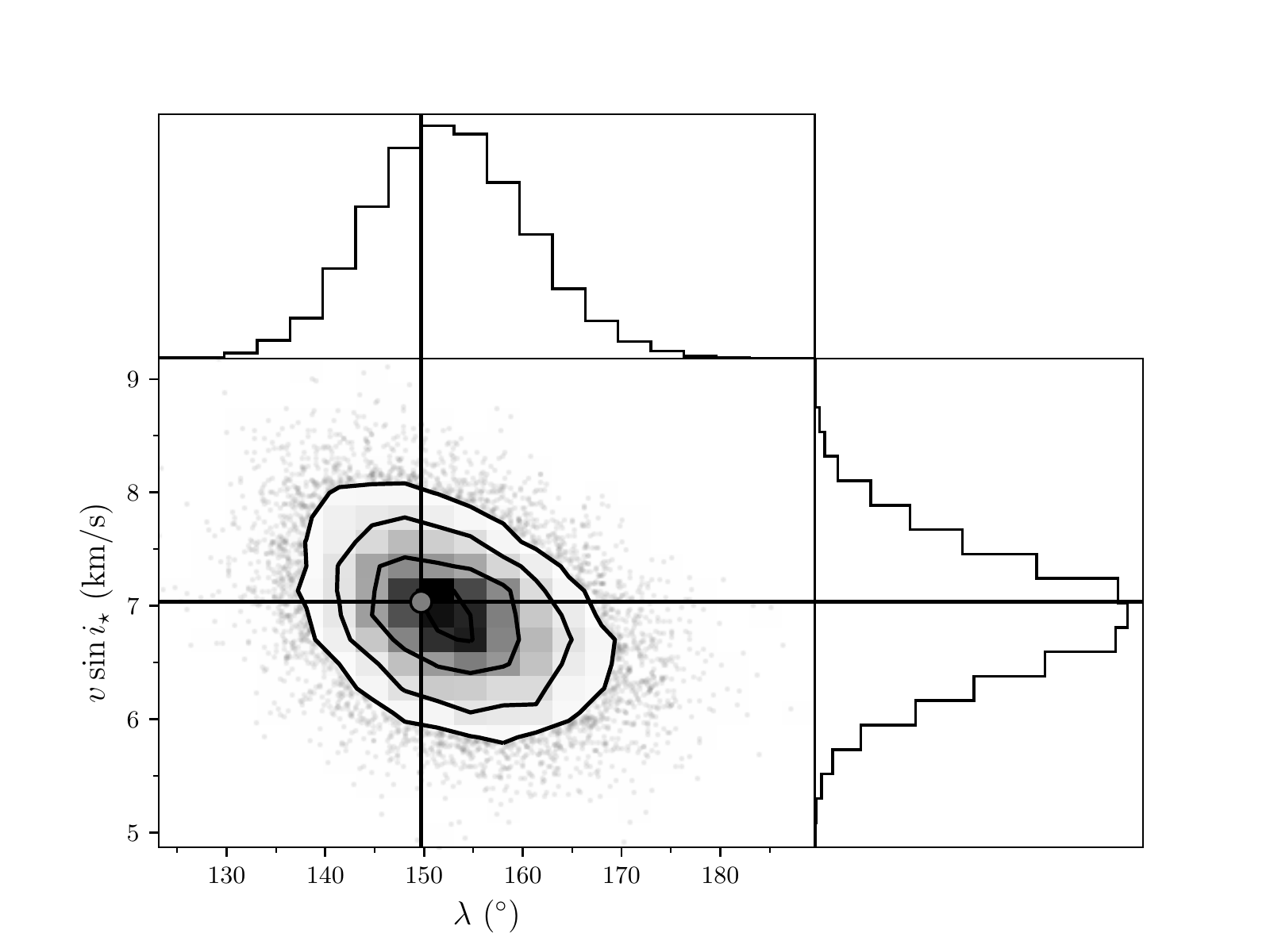}\includegraphics[width=0.5\textwidth]{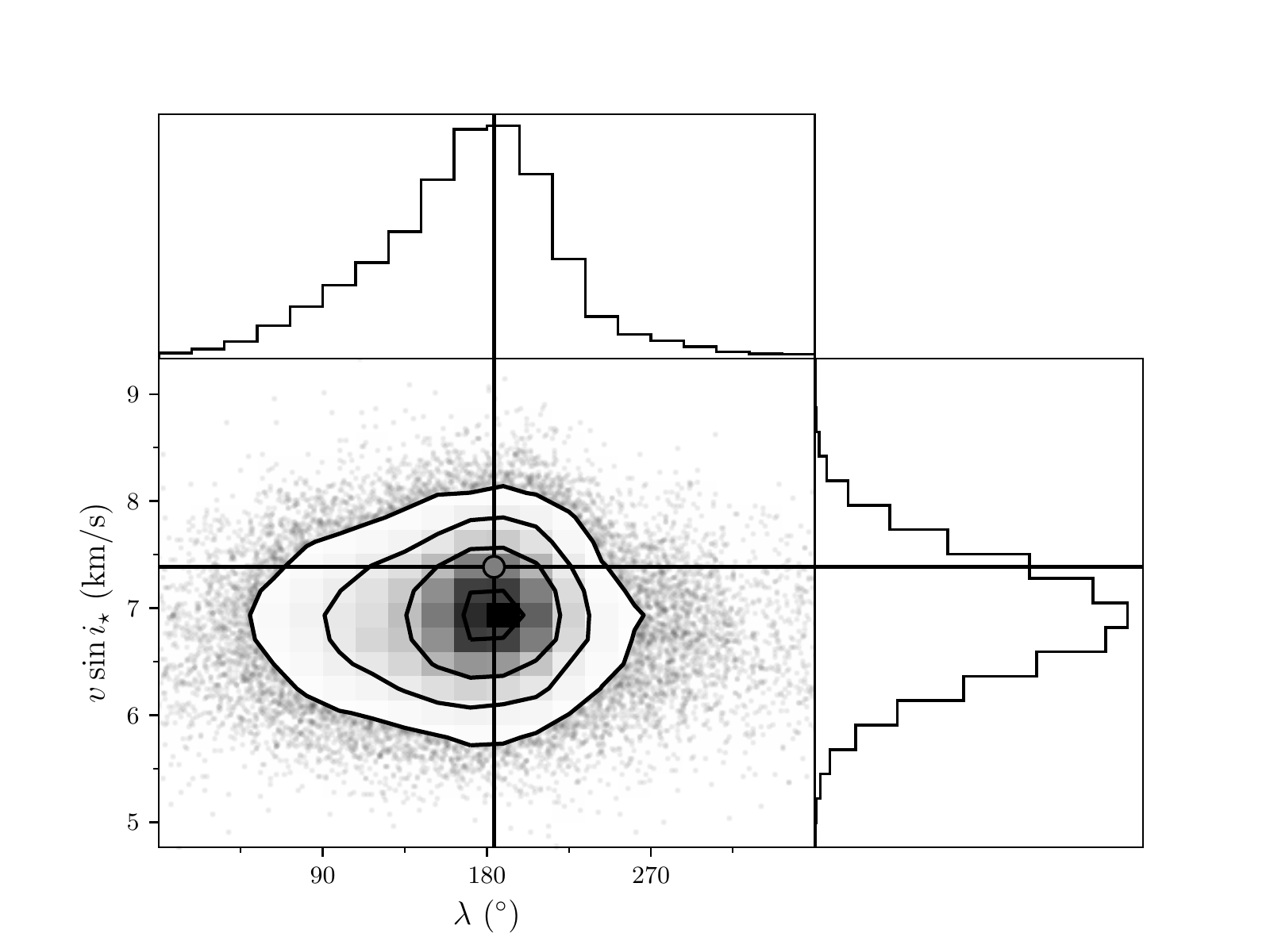} 
\caption{\label{fig:S2} {\bf Posteriors of $v\sin i_\star$, $\lambda_c$, and $\lambda_b$.} Left: Posterior for the projected obliquity with respect to planet c plotted against the projected stellar rotation. The gray dot denote the maximum likelihood values used to depict the models in Fig.~1 and Fig.~S1. Right: Same, but for planet b. Note that the $v \sin i_\star$ posterior from the run for planet c was used as a prior for planet b.}
\end{figure}

In Tab.~\ref{table:system_parameters} we report the results for all parameters based on the 15.9\%, 50\%, and 84.1\% levels of the cumulative marginalized posterior distributions. Fig.~1 in the main article and Fig.~\ref{fig:S1} show the RV and phase folded photometry data as well as the maximum likelihood model. The two panels in Fig.~\ref{fig:S2} display the correlation between the projected obliquities and $v \sin i_\star$.

 We also tested the robustness of $\lambda_c$ against the orbital eccentricity of planet c. The authors of the discovery paper \cite{Hjorth+2019} prefer a circular solution for the orbit, based on the available RVs and photometry. For such a circular orbit we obtain $\lambda_c=153 \pm 8$~deg. If we use their eccentric orbital solution ($e=0.144 \pm 0.033$ and $\omega = 70 \pm 9$~deg), then we find $\lambda=156^{+6}_{-7}$~deg. Our measurement of the projected obliquity does not strongly depend on the assumed orbital shape. We follow here the discovery paper and adopt the circular solution.

\begin{figure}
\includegraphics[width=0.5\textwidth]{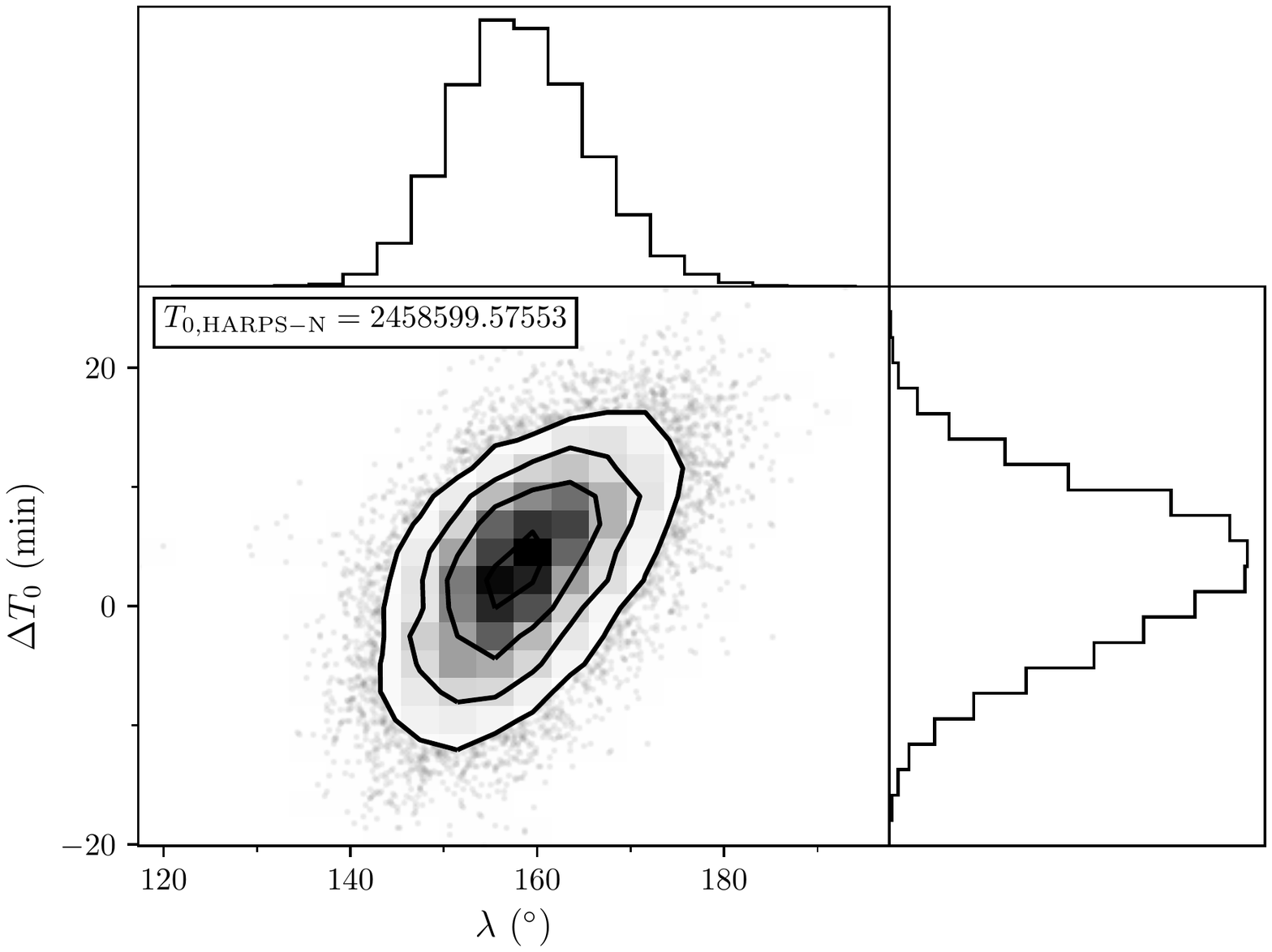} 
\caption{\label{fig:S3} {\bf Posteriors of $\Delta T0$ and $\lambda_c$.} Corner plot for the particular mid transit timing uncertainty for planet c during the night of our HARPS-N observation and for $\lambda_c$. For this transit event we found a posterior of $T_{\rm 0, HARPS-N} = 2458599.578 \pm 0.005$~BJD. For creating this plot we subtracted from the y-axis the mean value and expressed the time axis in minutes. }
\end{figure}

The {\it K2} mission only observed two transits of planet c and while we use information from out of transit RVs covering a time period of more than 500 days, RVs are not as powerful in constraining transit timing as transit photometry. To what extend does the uncertainty in $T_0$ and $P_c$ and consequently the uncertainty in the particular mid-transit times of our spectroscopic observations with HARPS-N and HDS affect $\lambda_c$? From the posterior of our MCMC run we find the uncertainty interval for the particular mid transit timing of our HARPS-N observations to be enlarged to $\sim$10~min, $T0_{\rm HARPS-N} = 2458599.578\pm0.005$~BJD. This can be compared to the $T_0$ interval listed in Tab.~\ref{table:system_parameters} ($2458019.1723\pm0.0003$~BJD), which reports the timing of twelve epochs earlier and closer to the {\it K2} observations. The HDS transit observations had been carried out exactly one epoch after the HARPS-N observations  and the timing uncertainty for that particular transit is unchanged to the uncertainty for the HARPS-N transit. The corner plot in Fig.~\ref{fig:S3} illustrates that this timing uncertainty does lead to an enlarged uncertainty interval for $\lambda_c$ and is accounted for in our final result for $\lambda_c=153 \pm 8$~deg.

Finally we tested if the two RV data sets from HDS and HARPS-N give consistent results independent from each other. For this we performed two additional MCMC runs, both identical to the above mentioned run, but with differing RV data sets. In the first run we only included HDS RVs and we find $\lambda_c=144^{-8}_{+7}$~deg. In the second run we only included the HARPS-N RVs and find $\lambda=167^{-24}_{+21}$~deg. The two RV data sets obtained with two different instruments during two different transit events give consistent results. 

\section{Alternative RM analysis for planet b}
As a further check on our results on the projected obliquity of planet b we analyzed the spectra at the level of spectral-line distortions (the planet's ``Doppler shadow''), rather than the level of anomalous radial velocities \cite{Albrecht+2007,CollierCameron+2010,Albrecht+2013,Johnson+2014,Zhou+2016,Cegla+2016}. For this exercise, we used the CCFs generated by the standard ESPRESSO data reduction software. First, we created a mean out-of-transit CCF based on the pre-ingress data. Then, we subtracted the mean out-of-transit CCF from the CCF of each ESPRESSO spectrum. Fig.~\ref{fig:S5}, panel C, shows these CCF residuals. Any distortion of the CCFs during the transit is not obvious to the eye. To proceed, we used a stacking technique \cite{Johnson+2014,Hjorth+2019b}. For a given choice of $\lambda_{\rm b}$ and $v\sin i_{\star}$, we identified those portions of the CCF residuals that should have been affected by the planetary shadow. These residuals were added -- weighting them by a function representing the local stellar surface velocity field governed by macro-turbulence -- to form a statistic that should be largest when $\lambda_{\rm b}$ and $v\sin i_{\star}$ are chosen correctly. We further weighted this statistic by a Gaussian prior on $v\sin i_\star$ with a mean of $6.9$~km~s$^{-1}$ and a standard deviation of $0.6$~km~s$^{-1}$, based on the analysis of the RM effect for planet c. Fig.~\ref{fig:S5}, panel F, shows the strength of this statistic as a function of $\lambda_{\rm b}$ and $v\sin i_{\star}$. 

\begin{figure*}
\centering
\includegraphics[width=1\textwidth]{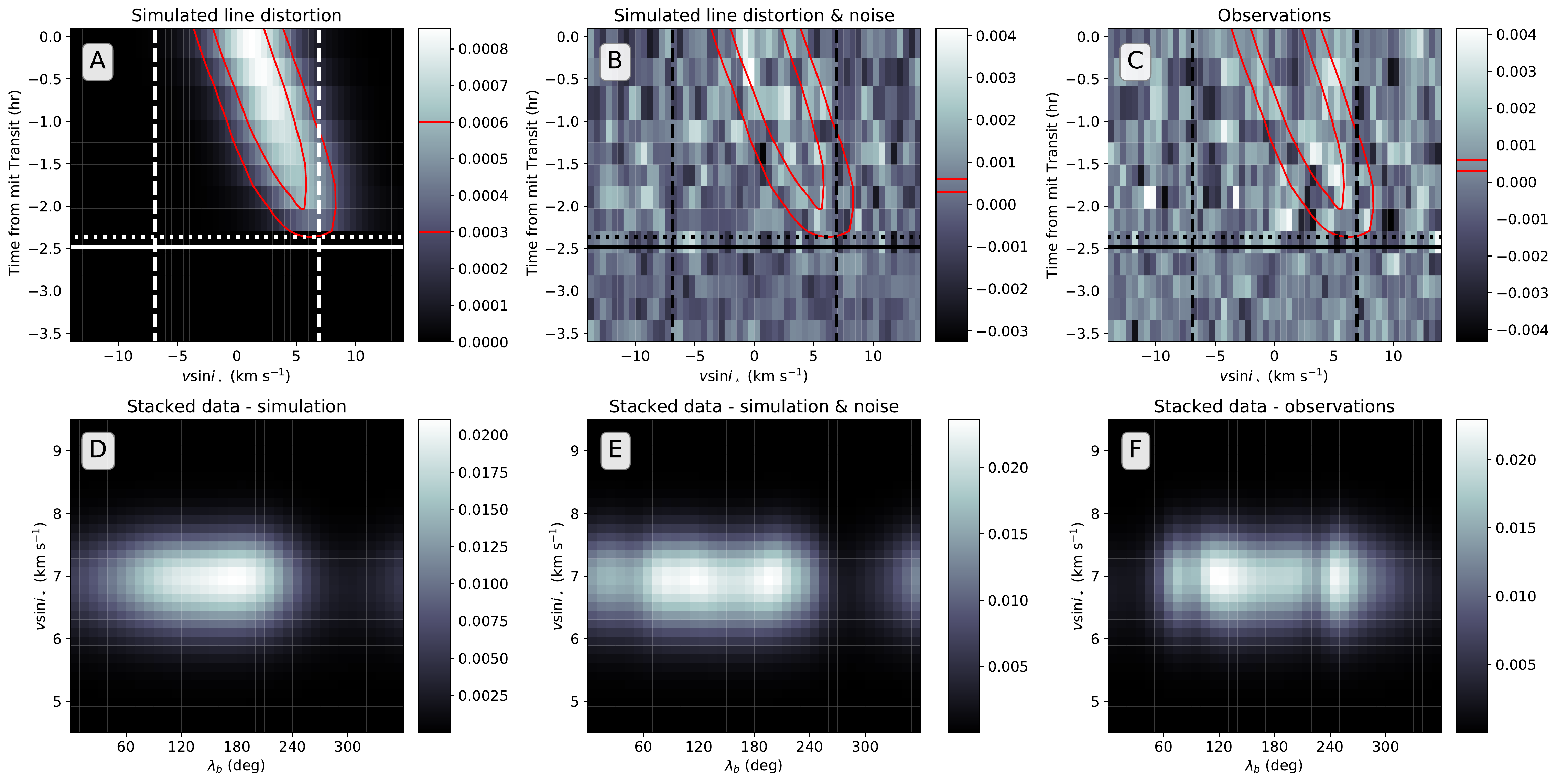} 
\caption{\label{fig:S4}{\bf The Doppler shadow of K2-290~A\,b in the ESPRESSO data.} Panel A: Noiseless simulation of the distortion in the cross-correlation functions based on the parameters of the best-fitting model to the light curve and the ESPRESSO RV data (Table~S1). The vertical dashed lines mark $v \sin i_{\star}$, while the horizontal solid and dotted lines mark the times of first and second contact. B: Same, after adding Gaussian random noise with an amplitude set equal to the observed noise level. C: Real data. The red outlines in panels A-C mark the positions of particular intensity levels in the simulation. These levels are also indicated in the color bars. The scales are calibrated so that the out of transit cross-correlation function has a height of unity. Panels D-F display the strength of the ``stacked signal'' for different choices of $v \sin i_\star$ and $\lambda_b$ in the simulation (D), the simulation with noise (E), and the real data (F).}
\end{figure*}

\begin{figure}
\includegraphics[width=0.5\textwidth]{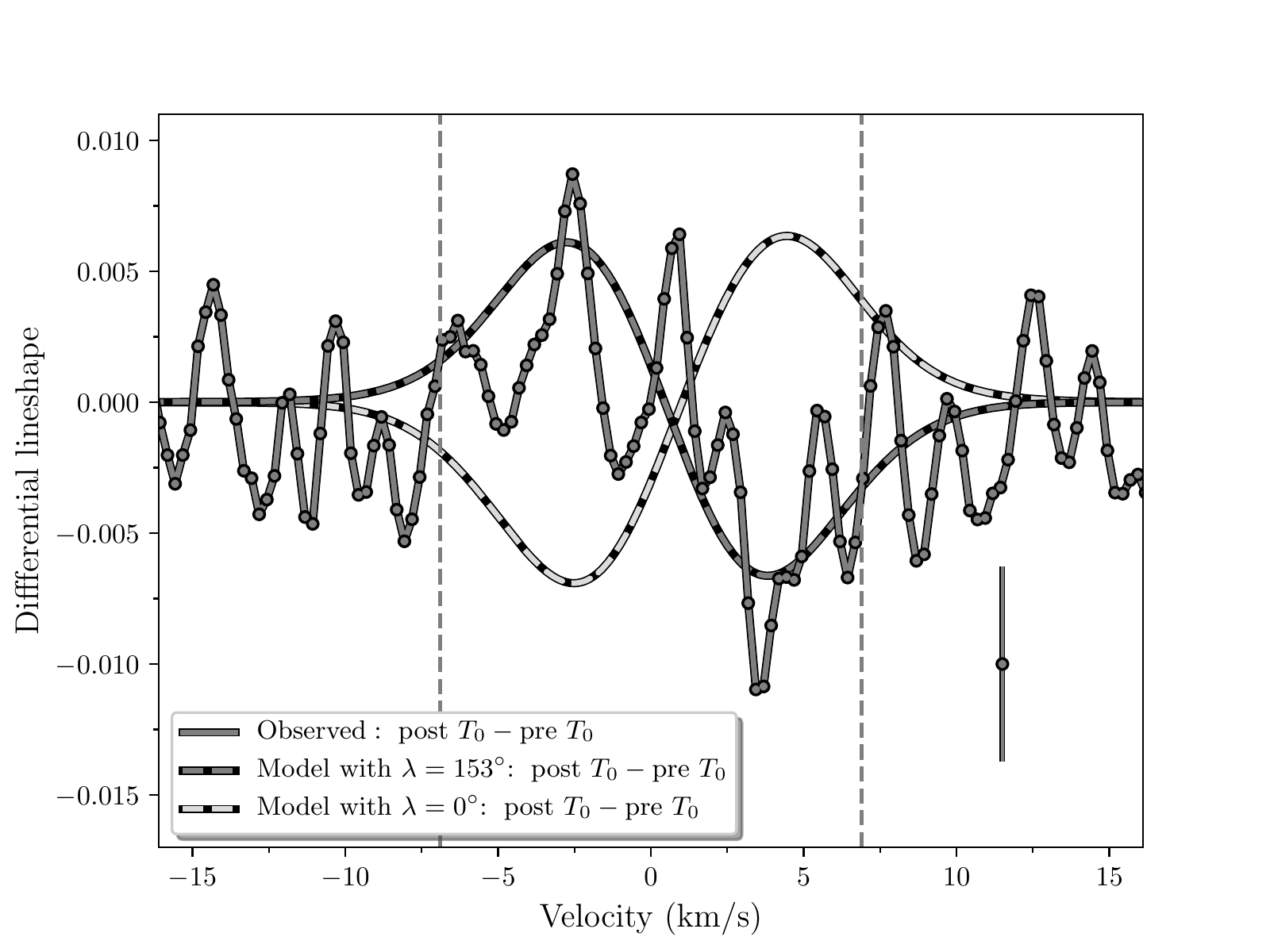} 
\caption{\label{fig:S5} {\bf Illustration of the line deformations in the HARPS-N data taken during transit of planet c.} The solid line with dots indicates the average post mid-transit CCF from which the mean pre mid-transit CCF was subtracted. The black-dark gray dashed line indicates the model perdition based on the best fitting values from a fit to the photometry and RVs ($\lambda_c = 153$~deg). The back-light gray dashed line shows a similar model with the difference that a well aligned orbit ($\lambda_c = 0$~deg) is assumed, a poor representation of the data. We note that the DRS (Data Reduction Software) of HARPS-N delivers CCF datapoints which are separated by 0.25~km\,s$^{-1}$ in velocity space. Given the spectral resolution of HARPS-N, R $\approx 110\,000$, the FWHM (Full Width Half Maximum) of the  PSF (Point Spread Function) is $\approx 2.72$~km\,s$^{-1}$. This leads to an oversampling of the CCF and therefore a "smooth" appearance of the noise. Adjacent datapoints are not completely independent from each other.}
\end{figure}

To help interpret the results, we created simulated data affected by the Doppler shadow, and then analyzed the simulated data with the same procedures used on the real data. An important parameter for the simulation is the macro-turbulent velocity: the stronger the macro-turbulence, the more the planetary-induced signal is smeared out in wavelength \cite{Albrecht+2012}. (Micro-turbulence and instrumental broadening also contribute to the smearing, but for an F star such as K2-290\,A, macro-turbulence is expected to be the largest contributor.) Our combined analysis of the real data gave an estimate for the macro-turbulent velocity of $4.8\pm 1.4$~km~s$^{-1}$. This is consistent with the value that was obtained by analysing the RM effect for another star of a similar effective temperature \cite{Albrecht+2013}. The scaling relations given in Refs.~\cite{Doyle+2014, Brewer+2016} lead to estimates of $5.4$~km~s$^{-1}$ and $7.3$~km~s$^{-1}$ for K2-290\,A. Based on all this information, we chose $4.8$~km~s$^{-1}$ as the basis for our simulations. We constructed simulated CCFs for various choices of $\lambda$ and  $v\sin i_{\star}$, convolved them with a model for the spectrograph broadening function (Fig.~\ref{fig:S4}, panel A), and added Gaussian noise to match the signal-to-noise ratio of the data (Fig.~\ref{fig:S4}, panel B). Through this numerical experimentation, we concluded that the apparent absence of a signal in Fig.~\ref{fig:S4}, panel C, is consistent with our simulations of the shadow given the system parameters and the achieved SNR in the observations. We also carried out the same stacking for our simulations with and without noise (Fig.~\ref{fig:S4}, panels D-E) and found the simulations with noise to resemble the weak detection of the RM signal shown in Fig.~\ref{fig:S4}, panel F.

We conducted a bootstrap analysis, creating simulated datasets by drawing randomly from the different CCFs, with repetitions allowed, and sampling form the posteriors of all the relevant parameters (period, mid-transit time, orbital inclination, scaled orbital separation, stellar projected rotation speed, radius ratio, and macro-turbulence). For the macro-turbulence we drew from a normal distribution with mean 4.8 and standard deviation 2~km\,s$^{-1}$. For each draw, we calculated the stacked signal for different values of $\lambda_b$. To this stacking signal as function of projected obliquity, we fitted a Gaussian function. We repeated this procedure 10{,}000 times. From the histogram we obtained $\lambda_b = 157\pm34^\circ$. Thus, the analysis of the Doppler shadow and the analysis of the anomalous RVs led to similar results. We place more confidence in the analysis of the anomalous RVs, because it is simpler than the analysis of the Doppler shadow.

\section{Alternative RM analysis for planet c}
Fig.~1 in the main article does demonstrate that the RM effect of planet c is clearly detected in RV space. However the plot also shows that no pre-ingress or post-egress data had been obtained during the transit nights. Can we confirm the detection of the RM effect using a different diagnostic? The HARPS-N spectra do not have iodine absorption lines imprinted. (The HDS spectra have been obtained through a iodine absorption line for wavelength calibration.) We therefore focus on the CCFs derived from the HARPS-N spectra by the Data Reduction Software of this instrument. We do not perform a classical analysis of the "Doppler shadow". The SNR in the CCFs is relatively low compared to the expected signal (Fig.~\ref{fig:S5}) and the observations had been interrupted around mid-transit. For these reasons we took the mean of the first 4 CCFs taken before mid-transit and subtract this mean from the average of the last 5 transit observations taken after mid-transit. (For a better estimation of the scale of the line deformations we have scaled all CCFs by the same factor so that they scale between [0-1]. This results in the same scaling as applied in Fig.~\ref{fig:S4} and the line deformations as caused by the two planets can be compared.) In Fig.~\ref{fig:S5} we show next to the data also the expected signal for our best fitting RV model with $\lambda_c = 153$~deg and an aligned model $\lambda_c = 0$~deg. The observed signal is fully consistent with the retrograde orbit. As for planet b we prefer the analysis of the RVs.

\section{Post-formation secular resonance scenario}

In the system's current configuration, planet b and c are too tightly coupled to the rotating primary star to experience significant misalignment by the M-dwarf. However, the system may have encountered a secular spin-orbit resonance after the disappearance of the gaseous disk. We investigated this possibility by numerically integrating the differential equations representing the coupled evolution of the orbital angular momentum of each planet, the orbital angular momentum of the binary orbit, and the stellar angular momentum. For details of this method, see \cite{Lai+2018}. The resonance requires a sufficiently low mass for planet b, no more than about $5~M_{\oplus}$, to avoid coupling the planets too strongly to the stellar spin. It also requires that the binary orbital distance is not much larger than about 100~au. We assumed the primary star has a moment-of-inertia constant of $k_\star = 0.06$ overall, and $k_q = 0.01$ for the rotational bulge \cite{Lai+2018}. We further assumed $M_{\rm b} = 5\,M_{\oplus}$, and adopted a somewhat arbitrary initial mutual inclination of $50$~deg for the binary orbit relative to the planetary orbital plane. The planets were initially aligned with each other and with the stellar spin. 

\begin{figure}
\centering
\includegraphics[width=\textwidth]{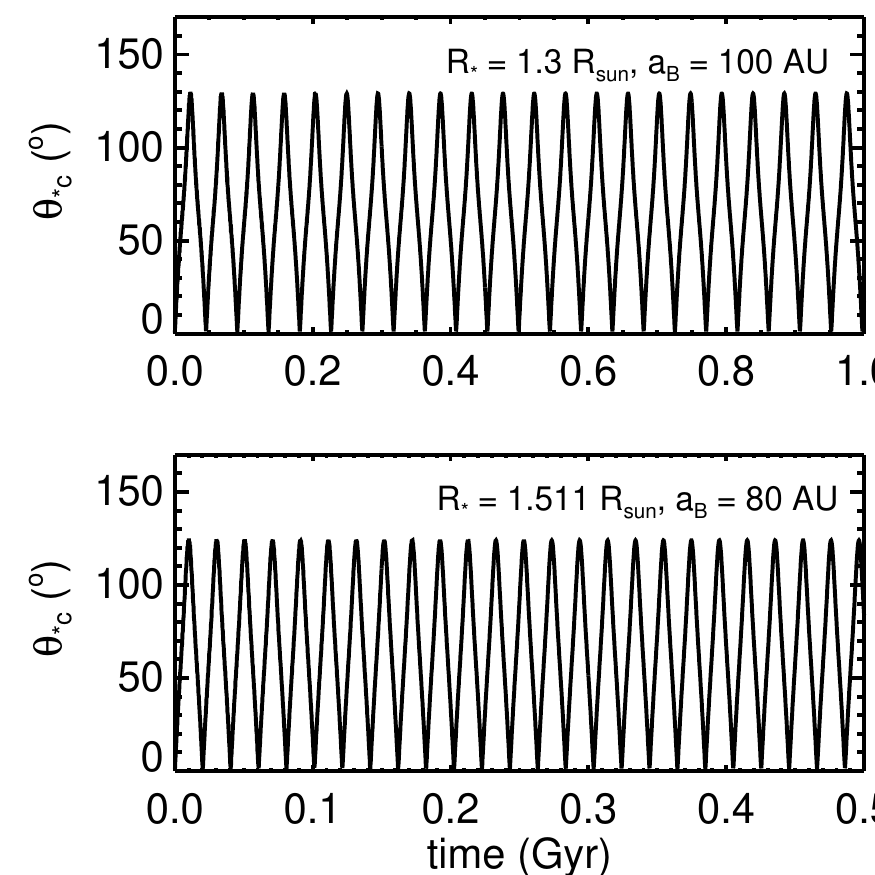} 
\caption{\label{fig:S6}{\bf Misalignment generated by secular resonance.}  Mutual inclination between the host star's equatorial plane and planet c's orbital plane ($\theta_{\rm sc}$) for two combinations  of  M-dwarf  separation and primary star radius.}
\end{figure}

\begin{figure}
\centering
\includegraphics[width=\textwidth]{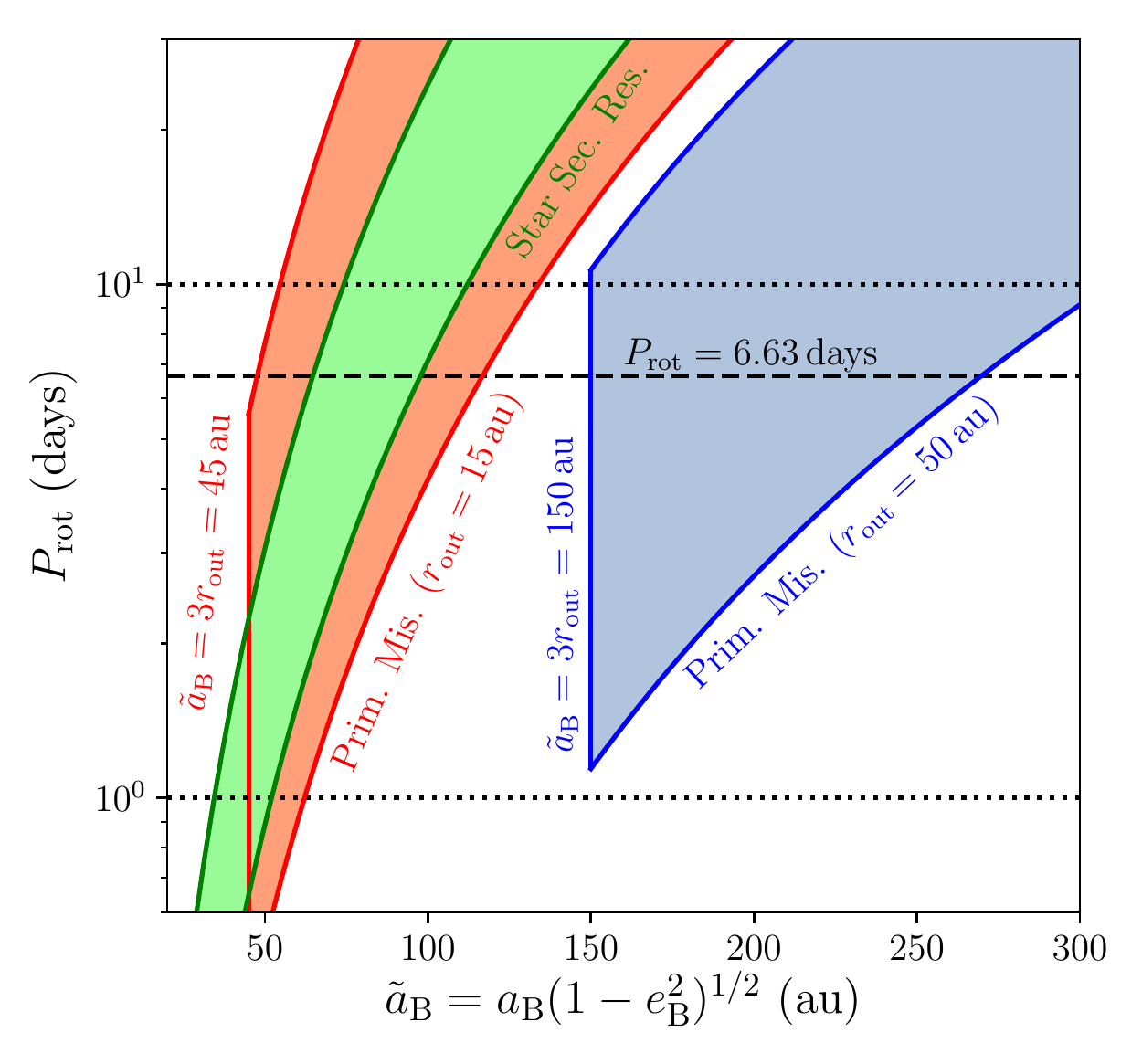} 
\caption{\label{fig:S7}{\bf Primordial and post-formation misalignment parameter space.}  Stellar rotation periods $P_{\rm rot}$ and re-scaled semi-major axis $\tilde a_{\rm B} = a_{\rm B}(1-e_{\rm B}^2)^{1/2}$, where $a_{\rm B}$ and $e_{\rm B}$ are the semi-major axis and orbital eccentricity of star B, respectively, capable of generating spin-orbit misalignment through secular resonances primordially ($\tilde \omega_{\rm dB} \sim \tilde \omega_{\rm sd} + \omega_{\rm sb} + \omega_{\rm sc}$), and after the gas disk dissipates ($\omega_{\rm cB} \sim \omega_{\rm sb} + \omega_{\rm sc}$, green shaded region).  Typical PMS stellar rotation rates ($1 \ {\rm day} \le P_{\rm rot} \le 10 \ {\rm days}$) are bracketed by dotted lines. The dashed line indicates the currently measured rotation period. The primordial misalignment shaded region is bracketed by $M_{\rm d} = 0.1 \ {\rm M}_\odot$ (upper line) and $M_{\rm d} = 0.005 \ {\rm M}_\odot$ (lower line) for $r_{\rm out} = 15 \ {\rm au}$ (red shaded region) and $r_{\rm out} = 50 \ {\rm au}$ (blue shaded region), where we require $r_{\rm out} \ge 3 \, \tilde a_{\rm B}$ from tidal truncation of the disk by star B. The lower limit for the disk mass ($M_{\rm d} = 0.005 \, {\rm M}_\odot \approx 5 \, {\rm M}_{\rm Jup}$) is consistent with the lowest gas-mass measurements of protostellar disks in the $\sigma$ Orionis Cluster \protect\citep{Ansdell(2017)}.} The post-formation shaded region is bracketed by $R_\star = 1.511 \ {\rm R}_\odot$ (upper line) and $R_\star = 1.0 \ {\rm R}_\odot$ (lower line).  We take $M_{\rm b} = 5 \ {\rm M}_\oplus$: all other parameter values are taken from primordial and post-formation misalignment scenario sections.
\end{figure}

In the first simulation (Fig.~\ref{fig:S6}, top panel), the star has not yet begun expanding and has an assumed radius of $1.3~R_{\odot}$, and the binary semi-major axis is  $a_{\rm B} = 100$~au. In the second simulation (Fig.~\ref{fig:S6}, bottom panel), the star has its current radius of $1.511 R_{\odot}$ and we set $a_{\rm B} = 80$~au, assuming a moderately eccentric orbit. In both cases, the planets remain strongly coupled and coplanar to each other, and develop a major misalignment relative to the stellar spin. The planets' strong gravitational coupling relies on planet c's large mass and the host star's moderate rotation period: a less-massive exterior planet ($M_{\rm c} \sim {\rm few} \ {\rm M}_\oplus$),  or a faster-rotating host star ($P_{\rm rot} \sim 1 \ {\rm day}$), would not enforce coplanarity under the influence of a misaligned stellar spin or companion  \citep{SpaldingBatygin(2016)}.  Based on
Eqn.~(16) of \cite{SpaldingBatygin(2016)}, the initial rotation period must have exceeded
$\approx$
$0.3 |i_{\rm b} - i_{\rm c}|^{-1/2}~{\rm day}$
for the two planets to have remained aligned at early times.  Moreover, if the masses of the planets were switched ($M_{\rm b} \sim {\rm M}_{\rm Jup}$, $M_{\rm c} \sim {\rm M}_\oplus$), secular resonances may destabilize the planetary system \cite{BatyginBodenheimerLaughlin2016}. This scenario does not require extreme fine tuning because the system could cross through the secular resonance when the gas disk disappears or as the stellar radius evolves. However these same choices of parameters would also have led to disk misalignment at earlier times (Fig.~\ref{fig:S7}).



\begin{table}
\caption{{\bf RV measurements taken during the transit of K2-290~Ac with HARPS-N@TNG}}
\begin{center}
\label{table:RVs_HARPSN}
\begin{tabular}{ccc}
\noalign{\smallskip}
\hline\hline
\noalign{\smallskip}
Time (BJD) & RV-19,760 (m~s$^{-1}$) & $\sigma_{RV}$ (m~s$^{-1}$)\\
\noalign{\smallskip}
\hline
\noalign{\smallskip}
2458599.511429 & -15.5 & 9.0 \\
2458599.525665 & -8.9 & 8.5 \\
2458599.539358 & -7.4 & 8.1 \\
2458599.553687 & -20.5 & 8.0 \\
2458599.665613 & 3.5 & 8.1 \\
2458599.680174 & 17.9 & 8.4 \\
2458599.694954 & 20.6 & 8.3 \\
2458599.708427 & -3.7 & 8.5 \\
2458599.734446 & 16.1 & 8.1 \\
2458599.755443 & -11.7 & 8.1 \\
2458653.461391 & -25.3 & 5.6 \\
2458662.439687 & -33.4 & 6.7
\end{tabular}
\end{center}
\end{table}

\clearpage

\begin{table}
\caption{{\bf RV measurements taken during the transit of K2-290~Ac with HDS@SUBARU.}}
\begin{center}
\label{table:RVs_HDS}
\resizebox{0.4\textwidth}{!}{%
\begin{tabular}{ccc}
\noalign{\smallskip}
\hline\hline
\noalign{\smallskip}
Time (BJD) & RV+12,000 (m~s$^{-1}$) & $\sigma_{RV}$ (m~s$^{-1}$)\\
\noalign{\smallskip}
\hline
\noalign{\smallskip}
2458646.904564 & 5.9 & 7.4 \\
2458646.915414 & 17.8 & 7.0 \\
2458646.926504 & 16.5 & 7.1 \\
2458647.750561$^{\star}$ & -7.2 & 6.3 \\
2458647.761661$^{\star}$ & -34.5 & 6.3 \\
2458647.773090$^{\star}$ & 31.1 & 9.7 \\
2458647.784190 & 9.1 & 6.7 \\
2458647.795279 & 1.5 & 6.8 \\
2458647.808349 & -13.4 & 6.5 \\
2458647.819449 & -18.6 & 6.4 \\
2458647.830548 & -18.0 & 6.4 \\
2458647.841638 & -21.5 & 6.3 \\
2458647.852737 & -12.8 & 6.2 \\
2458647.863827 & -28.9 & 6.1 \\
2458647.874926 & -18.8 & 5.9 \\
2458647.886016 & -22.2 & 6.5 \\
2458647.897115 & -25.7 & 6.6 \\
2458647.908215 & -20.5 & 6.8 \\
2458647.921004 & -14.5 & 6.7 \\
2458647.932104 & -21.7 & 6.9 \\
2458647.943204 & 2.6 & 6.9 \\
2458647.954303 & -11.2 & 6.9 \\
2458647.967083 & 2.2 & 6.7 \\
2458647.981662 & -0.5 & 6.3 \\
2458647.996231 & 15.1 & 6.7 \\
2458648.009101 & 5.6 & 6.8 \\
2458648.020200 & 16.7 & 7.4
\end{tabular}
}
\end{center}
\textit{Notes:} $^{\star}$Excluded in the analysis due to them being affected by twilight or technical problems with the CCD read-out.
\end{table}

\clearpage

\begin{table}
\caption{{\bf RV measurements taken during the transit of K2-290~A\,b obtained from ESPRESSO@VLT.}} 
\begin{center}
\label{table:RVs_ESPRESSO}
\begin{tabular}{ccc}
\noalign{\smallskip}
\hline\hline
\noalign{\smallskip}
Time (BJD) & RV-19,670 (m~s$^{-1}$) & $\sigma_{RV}$ (m~s$^{-1}$)\\
\noalign{\smallskip}
\hline
\noalign{\smallskip}
2458685.496193 & -2.6 & 1.9 \\
2458685.507030 & -2.4 & 1.9 \\
2458685.517732 & 2.4 & 2.0 \\
2458685.528853 & -1.6 & 2.0 \\
2458685.539639 & 2.8 & 2.6 \\
2458685.550695 & -0.1 & 2.6 \\
2458685.561700 & -2.0 & 2.6 \\
2458685.572653 & -5.1 & 2.1 \\
2458685.583376 & -3.5 & 2.1 \\
2458685.594217 & -1.8 & 2.1 \\
2458685.605165 & -3.5 & 2.4 \\
2458685.621852 & -3.9 & 1.8 \\
2458685.635662 & 0.1 & 1.9 \\
2458685.650111 & 2.3 & 2.3 \\
2458685.664976$^{\star}$ & -4.4 & 2.5 \\
2458685.679009$^{\star}$ & -6.1 & 2.2 \\
2458685.701177$^{\star}$ & -4.8 & 3.0 \\
2458685.728771$^{\star}$ & -5.7 & 2.1 \\
2458685.739713$^{\star}$ & -12.8 & 2.3
\end{tabular}
\end{center}
\textit{Notes:} $^{\star}$Excluded in the analysis due to the cloud coverage.
\end{table}

\begin{table}
\caption{\bf K2-290\,A system parameters obtained from transit data.} 
We report median and 1-$\sigma$ values. The ``Prior'' columns specify the distributions from which we draw our samples from; $\mathcal{U}$ denotes a uniform distribution, $\mathcal{N}(\mu,\sigma)$ and $\mathcal{T}(\mu,\sigma,a,b)$ denote a normal and truncated normal distribution, respectively, with mean $\mu$, standard deviation $\sigma$, lower boundary $a$, and upper boundary $b$.
\begin{center}
\label{table:system_parameters}
\resizebox{1.0\textwidth}{!}{%
\begin{tabular}{lcccc}
\hline\hline
\noalign{\smallskip}
Parameter & Value & Prior & Value & Prior \\
& \multicolumn{2}{c}{Planet c} & \multicolumn{2}{c}{Planet b} \\
\noalign{\smallskip}
\hline
Proj. spin-orbit angle $\lambda$ (deg) & $153 \pm 8$ & $\mathcal{U}$ & $173^{+45}_{-53}$ & $\mathcal{U}$ \\
Proj. stellar velocity $v\sin i_{\star}$ (km s$^{-1}$) & $6.9^{+0.5}_{-0.6}$ & $\mathcal{N}(\mu=6.5,\sigma=1.0)$ & $6.9 \pm 0.5$ & $\mathcal{N}(\mu=6.9,\sigma=0.6)$ \\
Orbital period $P$ (days) & $48.3674 \pm 0.0003$ & $\mathcal{U}$ & $9.2117 \pm 0.0002$ & $\mathcal{U}$ \\
Time of mid-transit $T_0$ (BJD) & $2458019.1723 \pm 0.0003$ & $\mathcal{U}$ & $2457994.7721^{+0.0014}_{-0.0015}$ & $\mathcal{U}$ \\
Scaled planetary radius $R_{\textrm{p}}/R_{\star}$ & $0.0684 \pm 0.0004$ & $\mathcal{U}$ & $0.0196 \pm 0.0003$ & $\mathcal{U}$ \\
Scaled orbital distance $a/R_{\star}$ & $42.9 \pm 0.9$ & $\mathcal{U}$ & $13.1 \pm 0.6$ & $\mathcal{U}$ \\
Orbital inclination $i$ (deg) & $89.34 \pm 0.06$ & $\mathcal{U}$ & $88.2^{+0.5}_{-0.6}$ & $\mathcal{U}$ \\
RV semi-amplitude $K_{\star}$ (m s$^{-1}$) & $38.5^{+1.7}_{-1.6}$ & $\mathcal{N}(\mu=38.4,\sigma=1.7)$ & $1.4^{+0.7}_{-1.4}$ & $\mathcal{T}(\mu=0.0,\sigma=2.2,a=0,b=\infty)$ \\
Systemic velocity HARPS-N $\gamma_{\textrm{HARPS-N}}$ (m s$^{-1}$) & $19,761^{+3}_{-4}$ & $\mathcal{U}$ & - & - \\
Systemic velocity HDS $\gamma_{\textrm{HDS}}$ (m s$^{-1}$) & $-11,996 \pm 3$ & $\mathcal{U}$ & - & - \\
Systemic velocity ESPRESSO $\gamma_{\textrm{ESPRESSO}}$ (m s$^{-1}$) & - & - & $19,699.1^{+0.8}_{-0.7}$ & $\mathcal{U}$ \\
Linear limb darkening coefficient in {\it Kepler}-band $u_{1,K}$ & $0.40 \pm 0.04 $ & $\mathcal{T}(\mu=0.38,\sigma=0.10,a=0,b=1)$ & $0.44 \pm 0.08$ & $\mathcal{T}(\mu=0.40,\sigma=0.04,a=0,b=1)$ \\
Quadratic limb darkening coefficient in {\it Kepler}-band $u_{2,K}$ & $0.10^{+0.04}_{-0.08}$ & $\mathcal{T}(\mu=0.14,\sigma=0.10,a=0,b=1)$ & $0.11^{+0.05}_{-0.06}$ & $\mathcal{T}(\mu=0.10,\sigma=0.06,a=0,b=1)$ \\
Linear limb darkening coefficient HARPS-N $u_{1,{\rm HARPS-N}}$ & $0.36 \pm 0.10$ & $\mathcal{T}(\mu=0.36,\sigma=0.10,a=0,b=1)$ & - & - \\
Quadratic limb darkening coefficient HARPS-N $u_{2,{\rm HARPS-N}}$ & $0.30 \pm 0.10$ & $\mathcal{T}(\mu=0.30,\sigma=0.10,a=0,b=1)$ & - & - \\
Linear limb darkening coefficient HDS $u_{1,{\rm HDS}}$ & $0.44^{+0.10}_{-0.09}$ & $\mathcal{T}(\mu=0.41,\sigma=0.10,a=0,b=1)$ & - & - \\
Quadratic limb darkening coefficient HDS $u_{2,{\rm HDS}}$ & $0.36^{+0.10}_{-0.09}$ & $\mathcal{T}(\mu=0.33,\sigma=0.10,a=0,b=1)$ & - & - \\
Linear limb darkening coefficient ESPRESSO $u_{1,{\rm ESPRESSO}}$ & - & - & $0.37 \pm 0.10$ & $\mathcal{T}(\mu=0.36,\sigma=0.10,a=0,b=1)$ \\
Quadratic limb darkening coefficient ESPRESSO $u_{2,{\rm ESPRESSO}}$ & - & - & $0.30 \pm 0.10$ & $\mathcal{T}(\mu=0.30,\sigma=0.10,a=0,b=1)$ \\

\end{tabular}
}
\end{center}
\end{table}

\FloatBarrier




